\documentclass{aa}
\usepackage{graphicx}
\usepackage{natbib}
\usepackage{txfonts}
\def\bp{\object{$\beta$\,Pictoris}}
\def\mic{\object{AU\,Mic}}
\def\vega{\object{Vega}}
\def\hd{\object{HD\,181327}}
\def\hr{\object{HR\,4796}}
\def\hdd{\object{HD\,139664}}
\def\hddd{\object{HD\,15115}}
\def\hdddd{\object{HD\,32297}}

\newcommand{\dma}[1]{_{\mathrm{#1}}}
\def\d{\mathrm d}
\begin{document}
\title{Collisional processes and size distribution in spatially extended debris discs}

\author{P. Th\'ebault\inst{1,2}, J.-C. Augereau\inst{3,4}}
\institute{
Stockholm Observatory, Albanova Universitetcentrum, SE-10691 Stockholm,
Sweden
\and
LESIA, Observatoire de Paris, Section de Meudon,
F-92195 Meudon Principal Cedex, France
\and 
Laboratoire d'Astrophysique de Grenoble (LAOG), Universit\'e Joseph Fourier,
B.P. 53, 38041 Grenoble Cedex 9, France
\and
Leiden Observatory, PO Box 9513, 2300 RA Leiden, The Netherlands}

\offprints{P. Th\'ebault} \mail{philippe.thebault@obspm.fr}
\date{Received; accepted. This version: \today} \titlerunning{collisional processes in debris discs}
\authorrunning{Th\'ebault and Augereau}

\abstract
%
{
New generations of instruments provide, or are about to provide, pan-chromatic images
of debris discs and photometric measurements, that requires new generations of models,
to in particular account for their collisional activity.
}
%
{
We present a new multi--annulus code for the study of collisionally
evolving extended debris discs.
We first wish to confirm and extend our early result obtained for a single--annulus
system, namely that the size distribution in realistic
debris discs always departs from the theoretical collisional ``equilibrium''
$\d N \propto R^{-3.5} \d R$ power law, especially in
the crucial size range of observable particles ($R\la 1\,$cm),
where it displays a characteristic wavy pattern. We also aim at studying how
debris discs density distributions, scattered light luminosity profiles,
and SEDs are affected by the coupled effect of collisions and
radial mixing due to radiation pressure affected small grains.
}
%
{
The size distribution evolution is modeled over 10 orders of magnitude,
going from $\mu$m-sized grains to $50$\,km-sized bodies. The model takes into
account the crucial influence of radiation pressure--affected small grains.
We consider the collisional evolution of a fiducial, idealized
$a$=$120$\,AU radius disc with an initial surface density $\Sigma(a) \propto a^{\alpha}$. 
Several key parameters are explored: surface density profile,
system's dynamical excitation, total dust mass, collision outcome prescriptions.
}
%
{
We show that the system's radial extension plays a crucial role and
that the waviness of the size distribution is amplified by inter--annuli interactions:
in most regions the collisional and size evolution of the dust is
imposed by small particles on eccentric or unbound orbits produced
further inside the disc.
Moreover, the spatial distribution of all
grains $\la 1\,$cm significantly departs from the initial profile
in $\Sigma(a) \propto a^{\alpha}$,
while the bigger objects, containing most of the system's mass, still
follow the initial distribution.
This has consequences on the scattered--light radial profiles which
get significantly flatter, and we propose an empirical law
to trace back the distribution of large unseen parent bodies
from the observed profiles.
We also show that the the waviness of the size distribution has
a clear observable signature in the far-infrared and at (sub-)millimeter
wavelengths. This suggests a test of our collision model, that requires
observations with future facilities such as Herschel, SOFIA, SCUBA-2 and ALMA.
We finally provide empirical formulae for the collisional
size distribution and collision timescale that can be used for future
debris disc modeling.
}
%
{}
\maketitle

\section{Introduction}

Extrasolar discs around young stars have been imaged for more than
two decades now. Observations and modeling have revealed
the great diversity of these systems, in particular regarding
the luminosity and density distribution of the dust component.
Systems with the lowest dust to star luminosity ratios have
been commonly labeled debris discs \citep[e.g.][]{lag00}.
The archetypal member of this group
is \bp, which has been thoroughly observed and modeled since
the first observation by \citet{smi84} \citep[see reviews by ][]
{vid94,kal95,arty97}. These systems are believed
to represent a later stage of disc evolution, where most
of the initial solid mass has already been accumulated into
planetary embryos or removed by collisional erosion and
pressure forces \citep[stellar radiation/wind pressure, see][ for recent reviews
on this subject]{grea05,mey06}. 
Simple order of magnitude estimates show that the dust in
these systems cannot be primordial and has to be constantly
replenished \citep{arty97}. Although cometary
evaporation could also be a possibility \citep{ligre98},
the most likely dust production mechanism is collisional erosion
of bigger solid objects \citep{arty97,dom03}.
This hypothesis is reinforced by the estimated ages of
these systems, which are generally more than $10^{7}\,$yrs old \citep{grea05}.
For such ages,
the standard planetary formation model \citep[e.g.][]{lis93} predicts
that most early stages of planetary formation, i.e. grain coagulation,
planetesimal formation, runaway and/or oligarchic accretion among
these planetesimals, should already be over and that these systems
should be made of large planetary embryos as well as smaller objects leftover
from the formation process. The presence of big embryos should
dynamically excite the system and lead to highly destructive
mutual encounters between the smaller leftover bodies \citep{ken04},
thus triggering a collisional cascade producing objects down to
very small dust grains.

The problems faced when modeling debris discs are numerous.
One first difficulty is that all objects
bigger than about $1\,$cm are completely undetectable by observations.
Current observations only probe the lower tail
of a collisional cascade among objects invisible to us. The challenge is
thus to reconstruct this hidden bigger object population from the observed
dust component. 
But even for particles in the ``observable'' range, it is very difficult
to get a coherent global picture. Each type of observations
(visible, near-IR, far-IR, mm,etc...) is indeed predominantly sensitive to
one particle size range and to one radial region
of the disc. And even when a large set of such observational data at
different wavelengths is available (including spatially resolved images,
as for example for $\beta$ Pictoris),
it does not allow to straightforwardly reconstruct the dust population.
This ``connecting the dots'' procedure is always model
dependent because it depends on many parameters,
linked to the dust's composition, temperature, optical properties
and size distribution, which can
never be unambiguously constrained in a non--degenerated way
(see for instance the thorough best--fit studies of \citet{ligre98}
and \citet{aug01} for $\beta$ Pictoris or \citet{su05}
for Vega).
One challenge is in particular to get a coherent link between
the mm--sized population, where most of the mass of the ``dust'' component
is supposed to lie but for which spatial information is usually very poor,
and the $\mu$m--sized grains, which should contain most of the 
optical depth and for which high--resolution observations are
more and more frequently obtained.

\section{Previous works and paper overview}
The most basic way to perform these reconstructions of the unseen big objects
population or to derive coherent models of the dust population
is to
assume that the classical collisional equilibrium size distribution of \citet{dohn69}
in $\d N \propto R^{-3.5}\d R$ holds for all object sizes $R$. However, there
are many reasons to believe that such an assumption is
probably misleading. As it has been shown by \citet[][hereafter TAB03]{theb03},
the main problem arises from the smallest grains, whose behaviour is strongly
affected by pressure forces imposed by the central star: radiation pressure
in the case of luminous stars, wind pressure for low-mass stars \citep[e.g.
\mic,][]{aug06,stru06}. For stars of mass $M_* \ga 1\,M_{\odot}$, one major
point is the presence of a minimum size cutoff $R\dma{PR}$, all objects $R<R\dma{PR}$
being blown away by radiation pressure. Qualitatively,
this depletion of $R<R\dma{PR}$ grains leads to an overdensity of slightly 
bigger grains $R_1$, because $R<R\dma{PR}$ grains are depleted and can no longer
efficiently destroy nor erode grains larger than $R\dma{PR}$. The overabundance
of $R_1$ grains, in turn, induces a depletion of $R_2$ objects with $R_2$ slightly
larger than $R_1$, etc... This domino effect propagates
towards bigger sizes and leaves a characteristic wavy size distribution,
with a pronounced succession of overdensities and depletions with respect
to the $R^{-3.5}$ power law \citep[e.g.][]{bag94,theb03,kriv06}. These discrepancies
with the $\d N \propto R^{-3.5}\d R$ distribution are reinforced by the fact that 
the smallest objects in the $R>R\dma{PR}$ range are put on very eccentric
orbits by radiation pressure and have a dynamical behaviour very different
from that of the bigger non radiation-pressure-affected bodies  
\citep[see][ for a thorough discussion on this topic]{theb03}.

In TAB03 we quantitatively studied these complex effects for the specific
case of the inner \bp\ disc. For this purpose, 
a statistical numerical code was developed, which quantitatively
follows the size distribution evolution of a population of solid bodies, in
a wide micron to kilometre size--range, taking into account the major effects
induced by radiation pressure on the smallest grains (size cutoff, perturbed dynamical
behaviour,...). Our main result was to identify an important departure 
from the $\d N \propto R^{-3.5}\d R$ law, especially in the $1\,\mu$m to $1$\,cm
range.
The main limitation of this code is that it considers a single, isolated annulus.
It can thus only be used to study a limited region at one given distance from the star
($\simeq 5\,$AU in the case considered) but not the system as a whole.
A multi--annulus approach is needed to achieve this goal. \citet{ken02,ken04}
have developed such statistical multi--annulus codes, which have been applied
to various contexts. These codes are in some respect more sophisticated
than the one used in TAB03, in particular because they follow the dynamical
evolution of the system (which is fixed in TAB03). Nevertheless, the price
to pay for following the dynamics is that the modeling of
the small grain population is very simplified, with all bodies below
a size $R \simeq\,1$\,m following an imposed $R^{-q}\d R$ size distribution,
thus implicitly overlooking the aforementioned consequences of the
specific behaviour of the smallest dust particles.
More recently, \citet{kriv06} used a different approach
based on the kinetic method of statistical physics. This model 
is able to follow the evolution of both physical size
and spatial distribution (1D) of a collisionally evolving idealized
debris disc, from planetesimals down to $\mu$m--sized grains. 
This model has also the added advantage of taking into account
a large range a unbound particles below the blow-out limit.
This innovative approach gave promising results for the specific
case of the \vega\ system. 
However, the modeling of collisional
outcomes is, as acknowledged by the authors themselves,
very simplified, with for instance all cratering impacts being neglected.

In this paper we present a newly developed multi--annulus
version of our code, aimed at studying the collisional evolution
of spatially extended systems. Intra and inter--annuli interactions,
due to the radial excursions of radiation--pressure affected small
grains, are considered. In addition to this new global scheme,
a new and improved modeling of collision outcomes is presented,
with particular attention paid to the crucial cratering regime
(Section \ref{sec:model} and Appendix).
In order to clearly identify and study the complex mechanisms
at play, we consider in the present study the case of a fiducial
idealized debris disc of $120$\,AU radial extension, and explore
surface density distributions in $\Sigma(a) \propto a^{\alpha}$
around the reference MMSN $\alpha = -1.5$ case, where $a$ is
the distance to the star.
The evolution of the system's size distribution, and
its significant departure from the standard Dohnanyi steady--state
power law, is followed until $t=10^{7}\,$yrs and is presented in
section \ref{sec:simu}. 
The role of several key free parameters, such as the system's
dynamical state, stellar mass and grain physical composition
are explored in section \ref{sec:param}.
The evolution of the system's spatial distribution, optical depth
and the correspondence between observed dust and unseen bigger parent
bodies is addressed in section \ref{sec:spat}.
In section \ref{sec:obs} we investigate the impact these results
have on important observables, in particular
the scattered light and thermal emission luminosity profiles as well as
the SEDs.
In section \ref{sec:discu} we discuss the robustness of our results
and derive empirical laws for the size distribution and collisional
lifetimes which might be extrapolated to any kind of
extended collisionally evolving debris disc. 
Conclusions and perspectives are presented in section \ref{sec:conclu}.
More specific studies of specific debris disc systems will be the purpose of a
forthcoming paper.

\section{Numerical model}
\label{sec:model}

\subsection{Structure}

Our code adopts the classical particle--in--a--box statistical approach
to follow the collisional evolution of a population of solid bodies
of sizes comprised between $R\dma{min}\leq\,R\dma{PR}$, where $R\dma{PR}$ is
the radiation pressure blow-out size, and $R\dma{max}\simeq 10$--$50$\,km.
The system is made of $N_a$ concentric annuli of width $\Delta a_{ia}$
and centered at distances $a_{ia}$ from the star.
Within each annulus $ia$, bodies are distributed into $n$
size bins, each bin corresponding to bodies of equal size $R_{i}$.
The evolution of the size distribution with time is given by
the estimated collision rates and outcomes between
all collisionally interacting $(ia,i)$ bins. 
For small particles produced in an annulus $ia$ and placed on
high-eccentricity or unbound orbits
by radiation pressure, collisions with bodies located within
all annuli crossed by their orbits are taken into account.
A detailed presentation of the model is given in Appendix\,\ref{sec:appA}.

One key parameter for our model, and any similar study for
that matter, is the prescription for the collision outcomes. 
We adopt the classical approach where
the outcome of an impact between a target of size $R_t$ and a projectile
of size $R_p$ depends on the ratio between the center of mass specific kinetic
energy of the colliding bodies $E\dma{col}$ and the the so--called
critical specific shattering energy $Q^{*}$, which depends on the
objects' sizes and composition. Depending on the respective values
of these two parameters, impacts result in catastrophic fragmentation,
cratering or accretion.
The collision--outcome prescription has been updated with respect to the one
in TAB03, in particular for what concerns the cratering regime. The new
model now also accounts for differential chemical composition within the
system, the main parameter being here the radial distance from the star
$a\dma{ice}$ below which water ice sublimates. The complete collision outcome procedure
is described in more details in Appendix\,\ref{sec:appB}. As explained in this
Appendix, we consider a ``nominal'' case for the fragmentation and cratering
prescriptions and with $a\dma{ice}=20\,$AU, but other cases are explored
(see section\,\ref{sec:collpres}).

The price to pay for following the size distribution over more
than 10 orders of magnitude in size is that we cannot accurately 
follow the dynamical evolution of the system, whose dynamical
characteristics have to be fixed as inputs. 
In this case, all CPU--time consuming calculations of mutual
impacting velocities and collision physical outcomes are performed once at
the beginning of the run \citep[e.g. TAB03,][]{kriv06}.
We shall therefore
implicitly assume that the disc has reached a quasi--steady dynamical
state, which holds for timescales longer than the ones considered in the
simulations. We consider identical average values for particle eccentricities
and inclinations for all size bins, with the exception of bins
corresponding to particles affected by radiation pressure for which
specific orbital characteristics are numerically derived (see Appendix).

For a more detailed description of our code, see the Appendixies
\ref{sec:appA} and \ref{sec:appB}.

\subsection{Initial conditions}
\label{sec:init}

As mentioned in previous sections, we consider here a fiducial idealized
debris disc, for which the initial spatial distribution follows
the classical Minimum Mass Solar Nebulae (MMSN)
profile derived by \citet{haya81}, where the surface number density 
is such that $\Sigma(a) \propto a^{-1.5}$, where $a$ is the distance from
the star. We consider a $11$ concentric annulus disc, that
extends from $a\dma{min}=10$\,AU to $a\dma{max}=120$\,AU, a typical range for
the radial extension of dusty debris discs.

The initial conditions are chosen in accordance with the current
understanding of debris discs, i.e. systems in which the bulk of
planetesimal accretion process is already over and large planetary
embryos are present. These large objects should dynamically excite the system,
and average eccentricities and inclinations in the disc may reach values
of the order of $0.1$ for Lunar--sized embryos \citep{arty97}.
We thus take $\langle e\rangle =0.1=2\,\langle i\rangle $
(with $\langle i\rangle$ in radians) as our nominal dynamical conditions
and explore different orbital values in separate runs.
We follow the collisional evolution of all objects in
the $[R\dma{min},R\dma{max}]$ range, where $R\dma{min}\leq R\dma{PR}$
and $R\dma{max}\simeq 50\,$km. We take as a reference value
$R\dma{PR}=5\,\mu$m, which corresponds to the value for a compact grain
around a \bp\ -like star (A5V), but other possible $R\dma{PR}$ values
for earlier and later type stars are also explored (section \ref{sec:radpre}).
The planetary embryos themselves are left out of our study since
they are too isolated to contribute to the continuous collisional cascade, and
can only affect the dust production rate through sudden isolated events \citep[for
the detailed study of such violent events, see][]{grig07}.
We shall assume that the initial size distribution at $t=0\,$yr follows the
classical $\d N \propto R^{-3.5}\d R$ power law from $R\dma{min}$ to $R\dma{max}$
and we follow subsequent departures from this ``equilibrium'' distribution
as time goes by.
\footnote{However, the initial size--distribution is not a crucial
parameter, since test runs have shown that the system always settles to
the same steady state regardless of the initial $dN/dR$ prescription.}
Having fixed the initial size--distribution, the initial disc mass
is a free parameter which is explored in separate runs. 
This disc mass is parameterized by $M\dma{dust}$,
the total amount of "dust", i.e. grains smaller than $\simeq 1\,$cm,
in the system. The parameter $M\dma{dust}$ has been chosen as a reference because it is
usually the most reliable constraint on the disc mass which can be
derived from observations, since larger objects are observationally
undetectable. Most of this dust mass is believed to be
contained in the bigger millimetre--sized grains detected
at sub--millimetre to millimetre wavelengths.
Such millimetre wavelength surveys have shown that for debris discs around young
main sequence stars, $M\dma{dust}$ is typically comprised 
between $0.001$ and a few $0.1\,M_{\oplus}$
\citep[e.g. review by][ and references therein]{grea05}.
Accordingly, we shall consider two limiting cases: a low mass disc
with $M\dma{dust} = 0.001 M_{\oplus}$, and a high mass disc with
$M\dma{dust} = 0.1 M_{\oplus}$
(in both cases, the initial distribution of bigger objects is
obtained by extrapolating a $\d N \propto R^{-3.5}\d R$ size distribution
up to $R\dma{max}$).
Particles within the size bins are assumed to be compact silicates
in the regions closer to the star than the subimation limit $a_{sub}$ 
and compact ices beyond $a_{sub}$,
with $a_{sub}\simeq 20$AU in the nominal case (see section B.1 of the Appendix).

For each run, we let the system evolve for $10^{7}\,$yrs.
Of course debris discs
can have ages exceeding by far this value (as for instance \object{Vega} or
\object{$\epsilon$ Eridani}), and longer timescales should in principle be considered
here.
We should however restrict ourselves to $10^{7}$yrs because in most of
the cases the system reaches a steady-state much earlier than this
(typically after $\sim 10^{6}$years for our nominal case). The only exception
to this is the ``low--mass'' case with $M\dma{dust}=0.001M_{\oplus}$, for which
the steady state is not reached, at $t\dma{final}=10^{7}$yrs, in the outer regions
of the systems. For this specific case, we let the system
evolve until $10^{8}$yrs.

All initial parameters for the nominal high mass case are summarized in
Table\,\ref{init}.

\begin{table}
\begin{center}
\caption[]{Nominal case setup. The fields marked by a $\surd$
are explored as free parameters in the simulations. See text for details.}
\label{init}
\begin{tabular}{lll}
\hline
& Radial extension & $10<a<120\,$AU\\
& Number of annuli $\times$ radial width  & $11\times 10$\,AU\\
$\surd$ & Initial surface density profile & $\Sigma(a) \propto a^{-1.5}$\\
$\surd$ & Total ``dust'' mass ($R<1\,$cm) & $0.1\,M_{\oplus}$\\
 & Size range modelled & $3\,\mu$m$<R<50$\,km\\
 & Number of size bins & $103$\\
 & Initial size distribution & $\d N \propto R^{-3.5} \d R$\\
$\surd$ & Sublimation distance (water ice)& $a\dma{sub}=20\,$AU\\
$\surd$ & Dynamical excitation & $\langle e\rangle =0.1=2\,\langle i\rangle$\\
$\surd$ & Stellar type & A5V\\
$\surd$ & Blow out size & $R\dma{PR}=5\,\mu$m\\
$\surd$ & Collision outcome prescription & (see Appendix \ref{sec:appB})\\
\hline
\end{tabular}
\end{center}
\end{table}

\section{Results for the nominal case}
\label{sec:simu}

\subsection{High--mass disc ($M\dma{dust} = 0.1\,M_{\oplus}$)}
\label{sec:nomin}

\begin{figure*}
\makebox[\textwidth]{
\includegraphics[width=\columnwidth]{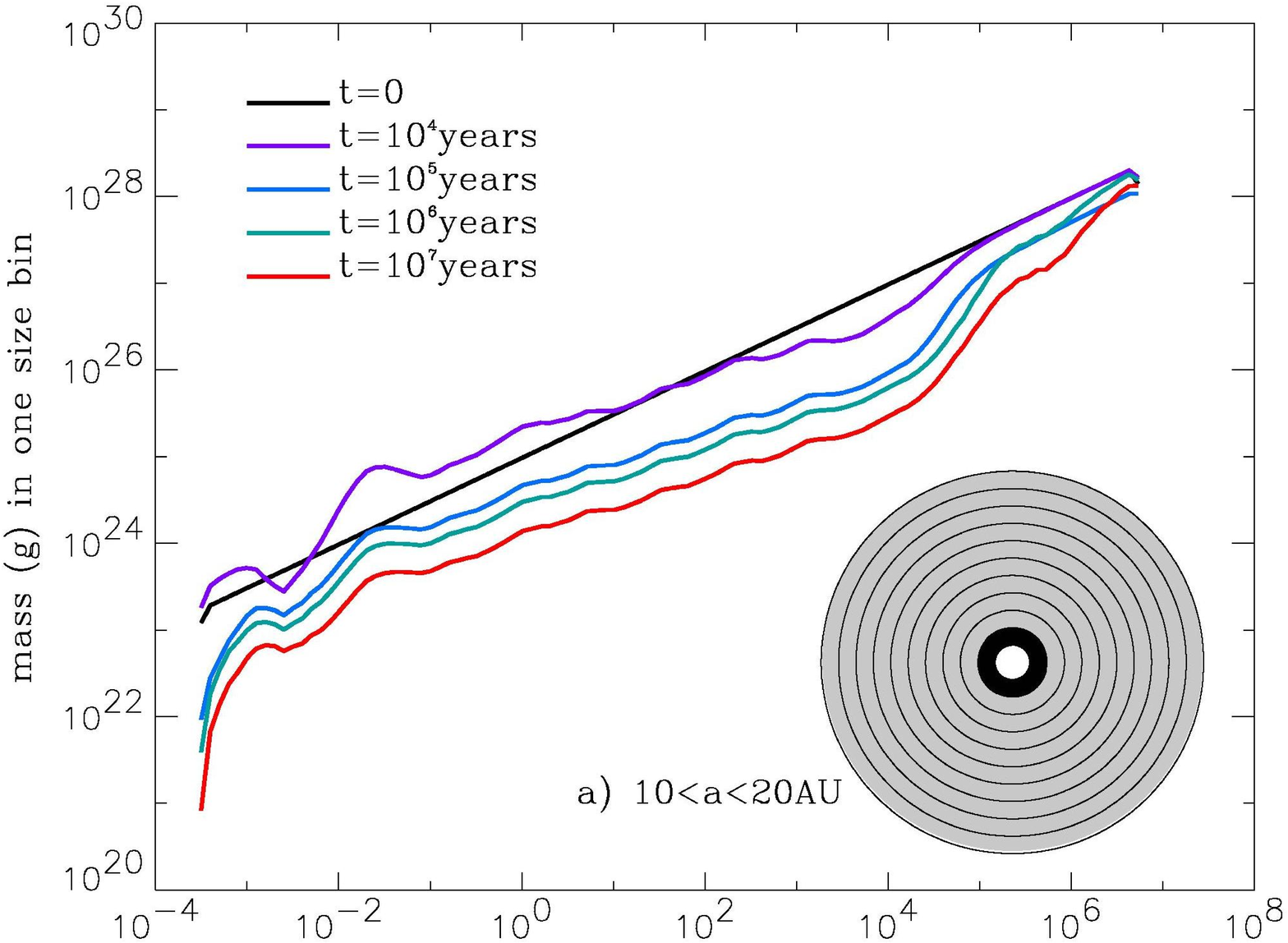}
\hfil
\includegraphics[width=\columnwidth]{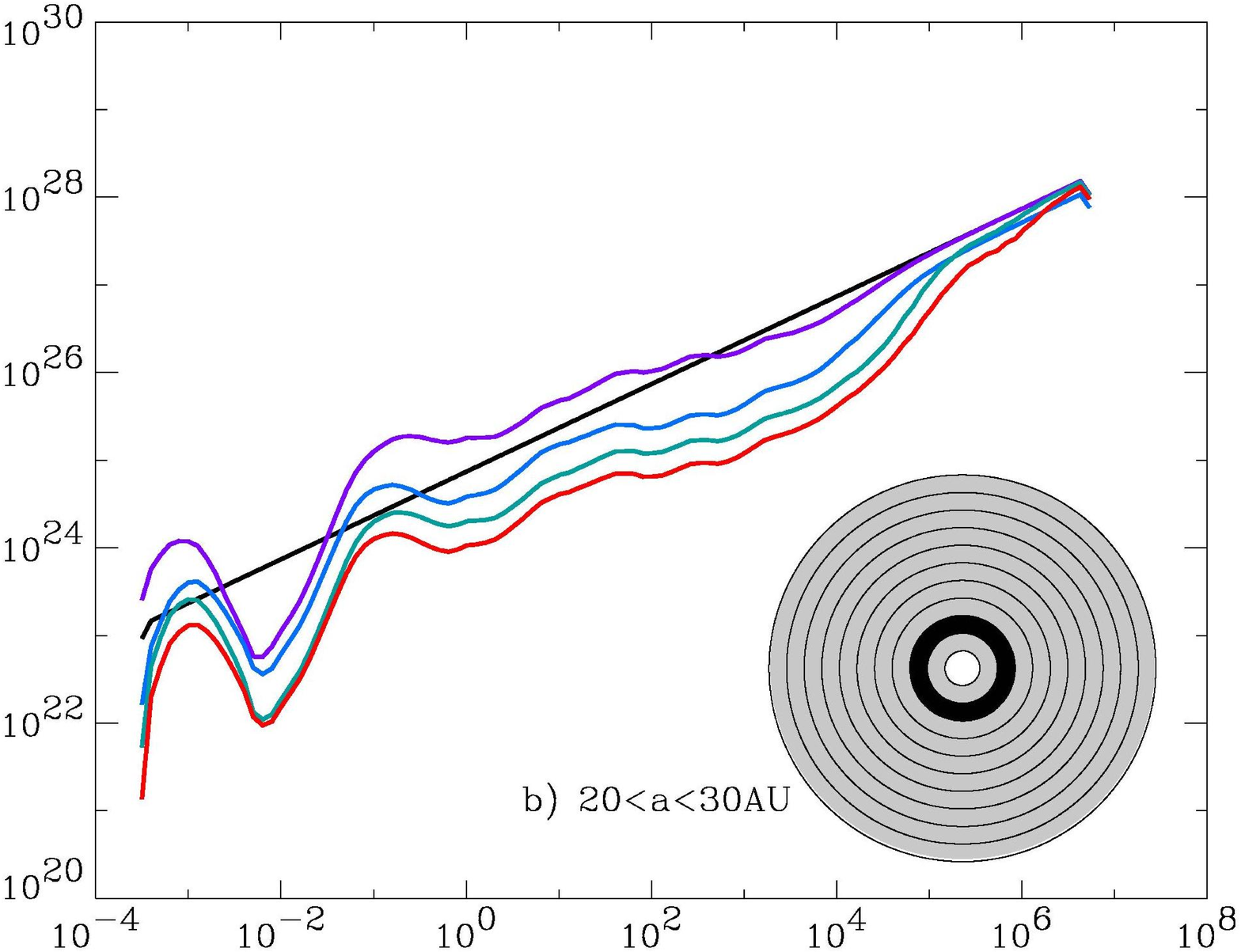}
}

\makebox[\textwidth]{
\includegraphics[width=\columnwidth]{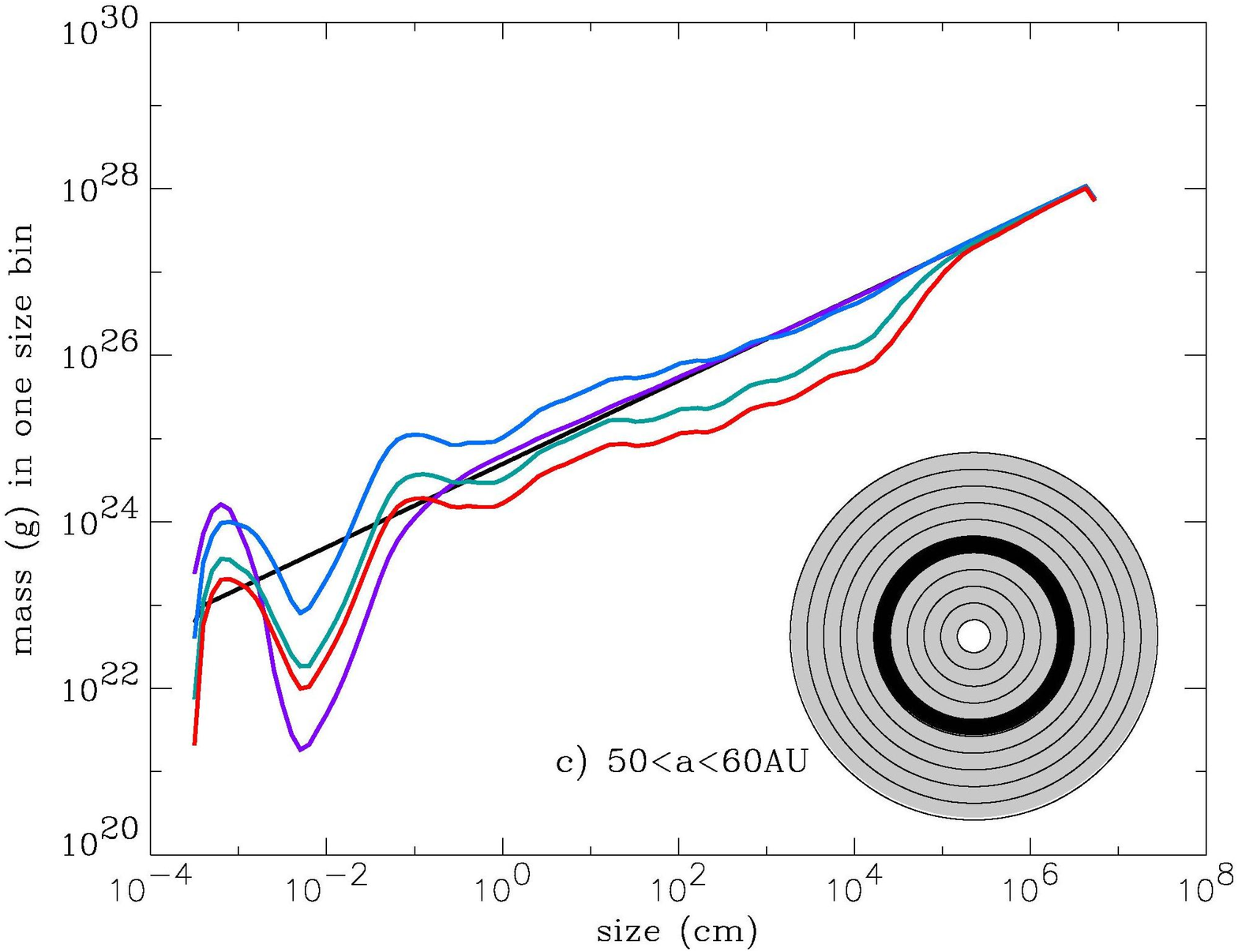}
\hfil
\includegraphics[width=\columnwidth]{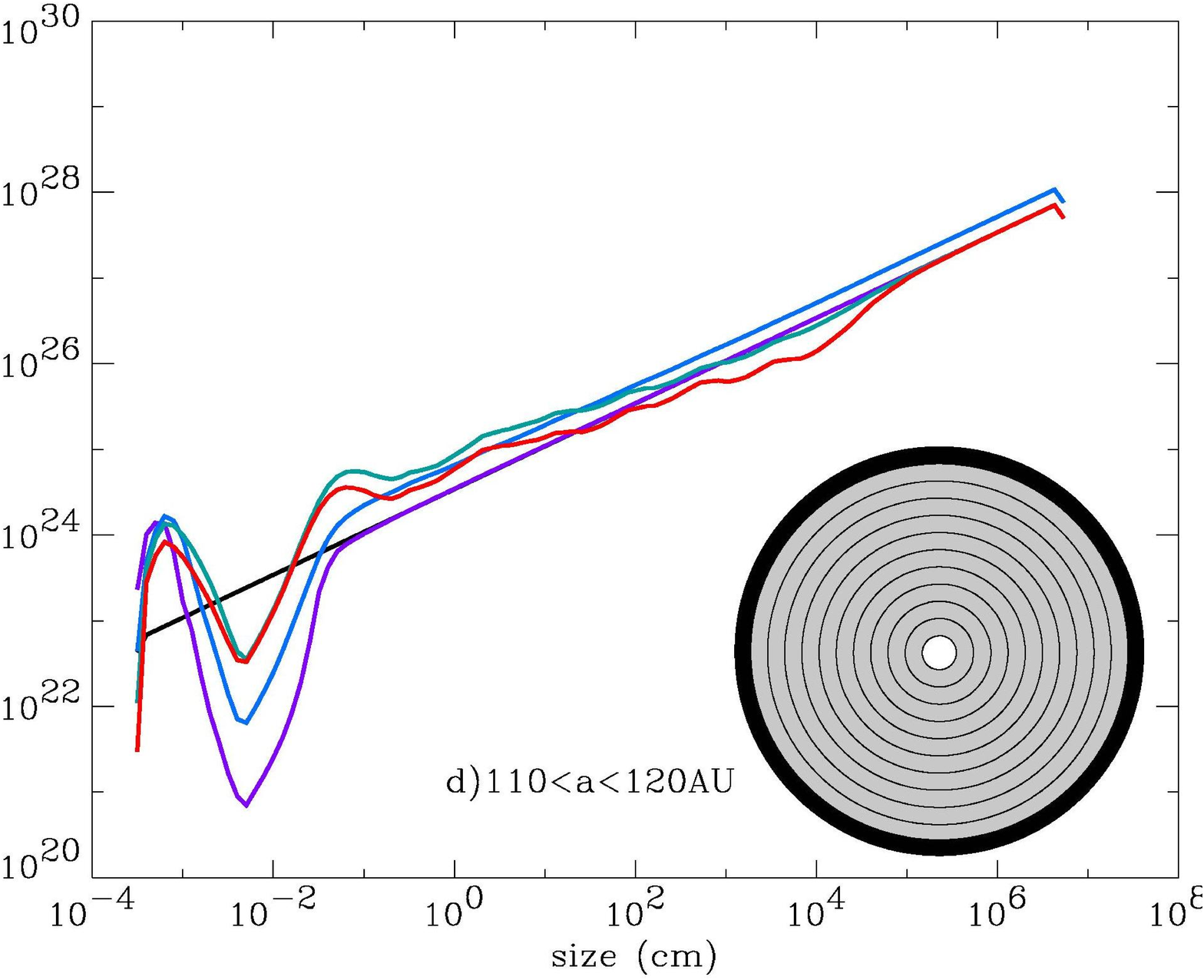}
}
\caption[]{High--mass case ($M\dma{dust}=0.1M_{\oplus}$).
Time-evolution of the size distribution at four different locations in the disc.
Note that the $y$-axis displays the mass contained in one size bin, which is
a correct way of displaying the mass distribution since all size bins are
equally spaced in a logarithmic scale.
}
\label{snapsh}
\end{figure*}

Figure~\ref{snapsh} shows the evolution of the size distribution for four
annuli at different distances from the star. In the innermost annulus
(Fig.\,\ref{snapsh}a), a weak wavy pattern develops, starting with the
depletion of $R<R\dma{PR}$ grains and propagating upward.
Once the pattern has fully developed, subsequent evolution consists in
a progressive total mass loss while the global size distribution
profile is conserved.
This wavy pattern is however much less pronounced in this
innermost annulus than in TAB03.
The first reason for this is that TAB03 considered a region
further inside the disc, at $5$\,AU, while the present inner annulus
starts at $a=10\,$AU and extends up to $20\,$AU. Impact velocities,
and their destructive efficiency, are thus significantly lower here.
The second reason is due to our revised collision--outcome prescription,
in particular for cratering events, which in TAB03 had a dominant role in
shaping the size distribution in the $R\la 1\,$cm domain (see Table 4
of this paper). With the more realistic
cratering prescription taken here, excavated masses are significantly smaller
in the small grains domain than in TAB03 (see Sec.\,\ref{sec:appB3}),
hence the shallower patterns in the size distribution.
The knee in the distribution around $0.1$--$1$\,km is a well known feature
\citep[e.g][]{bag94} due to the switch
from the strength dominated regime, where bodies resistance weakly decreases
with increasing size, to the gravity dominated regime, where bodies
resistance to impacts rapidly increases with increasing size.
It can be easily checked that the location of the knee at $R\simeq 0.1\,$km
corresponds to the least impact--resistant bodies (see Equ.\,\ref{Q}).
Furthermore, for large objects, reaccumulation of fragments after an impact
also begins to play a major role.

\begin{figure}
\includegraphics[angle=0,origin=br,width=\columnwidth]{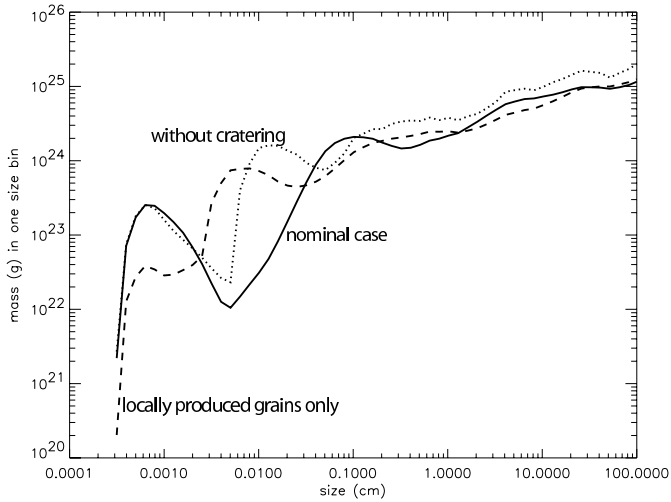}
\caption[]{Size distribution for the 50--60AU annulus, at $t=10^{7}\,$yrs,
for the nominal case (solid line),
when only taking into account fragmenting impacts,
i.e. no cratering (dotted line),
and when only taking into account the $locally$ produced grains,
i.e. no impact with grains coming from inner annuli (dashed line)
 }
\label{cutcra}
\end{figure}

In the more distant annuli, on the contrary, very pronounced wavy
patterns are observed in the size--distribution (Figs.\,\ref{snapsh}b,c and d).
The most striking features are the overdensity of $R\simeq 1.5\,R\dma{PR}$ bodies,
and above all, the strong depletion of bodies in the submillimetre range
($\simeq$ 10--50$\,R\dma{PR}$).
This result might appear counter--intuitive since
one would expect these features to be even less pronounced than
in the innermost annulus because of the
longer dynamical timescales and lower impact velocities in the outer
regions. The main cause for these sharp features are in fact small high--$\beta$
grains originating from other annuli further inside the disc
(where $\beta$ classically designates the radiation pressure to gravitation
forces ratio). This is clearly illustrated in Fig.\,\ref{cutcra} which
compares, in the middle $50$--$60$\,AU annulus, the final size distribution
(solid line) with the size distribution obtained when only
considering locally produced particles (dashed line).
In the $R \la 30\,\mu$m range,
foreign--born grains make up up to $90$\% of the local population, thus
resulting in a factor $\simeq 10$ increase of the number density. But
the effect of these additional inner--disc--produced grains on the
system's evolution exceeds by far that simply due to a number density increase
of an order of magnitude. Indeed, as these grains have had more time to reach high
radial velocities than the locally produced grains of the same size,
they will impact objects in the annulus at much higher relative velocities.
As an example, for a target on a circular orbit at $50$\,AU, an impact
by a locally produced small grain with $\beta=0.45$,
will occur at $\Delta v \simeq 1$\,km.s$^{-1}$, whereas an impact by a $\beta=0.45$
grain produced at $10\,$AU will occur at $\Delta v \simeq 5\,$km.s$^{-1}$.
This will result in much more destructive collisions.
It is this higher destructive efficiency which is responsible
for the deep depletion of objects up to $\simeq 100R_{PR}= 0.5$\,mm. 
Another important result is that a large fraction of the sub--mm grain
depletion is due to cratering impacts, as appears clearly
from the test run with no--cratering shown in Fig.\,\ref{cutcra} (dotted line).
Indeed, small $R<30\,\mu$m grains cannot directly break--up objects
bigger than $\simeq0.1\,$mm, even for their increased impact velocities,
while they can efficiently erode by cratering bodies up to
almost $\simeq1\,$cm.

\subsection{Low--mass disc ($M\dma{dust}=0.001\,M_{\oplus}$)}
\label{sec:lowm}

\begin{figure*}
\makebox[\textwidth]{
\includegraphics[width=\columnwidth]{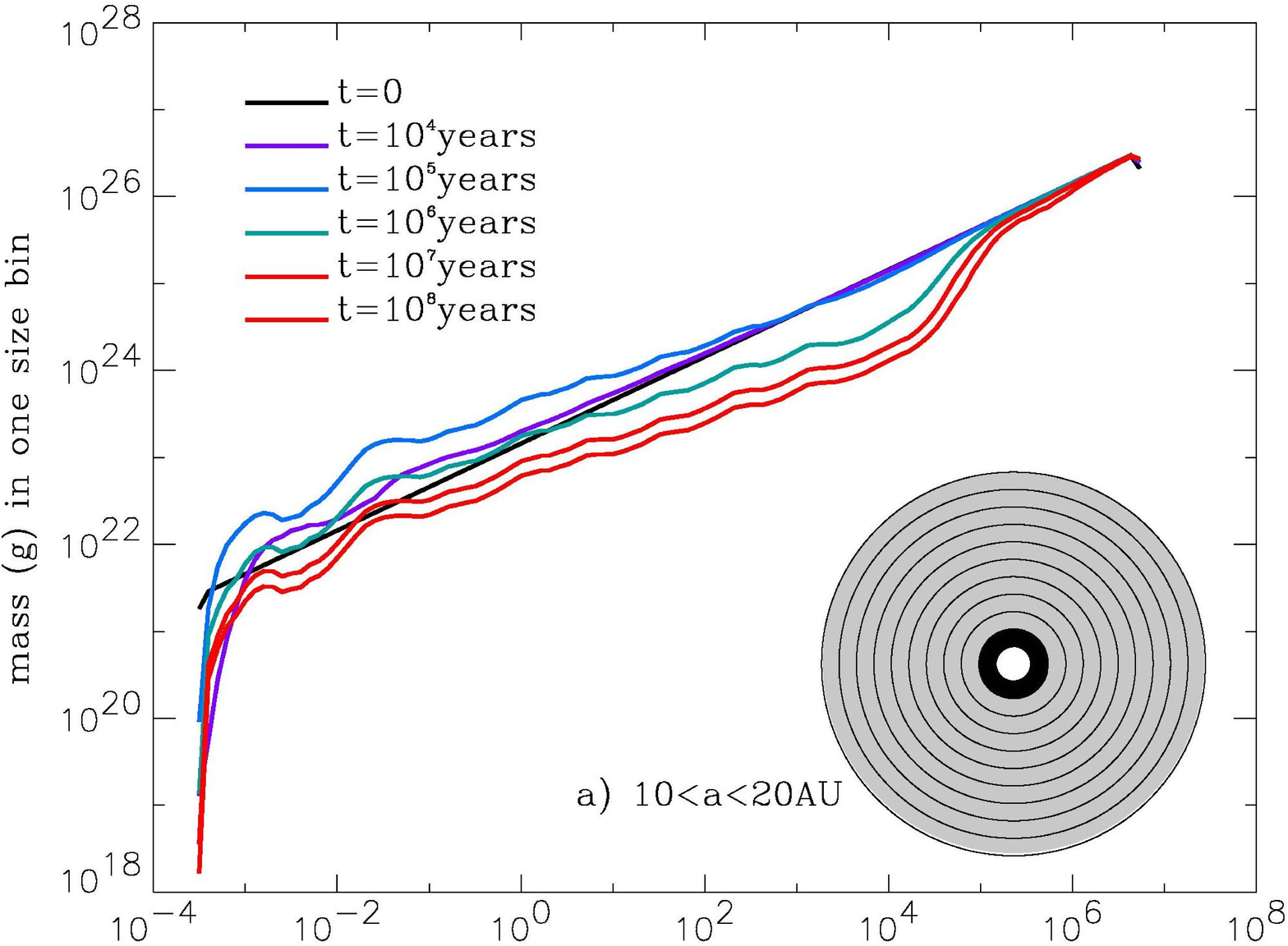}
\hfil
\includegraphics[width=\columnwidth]{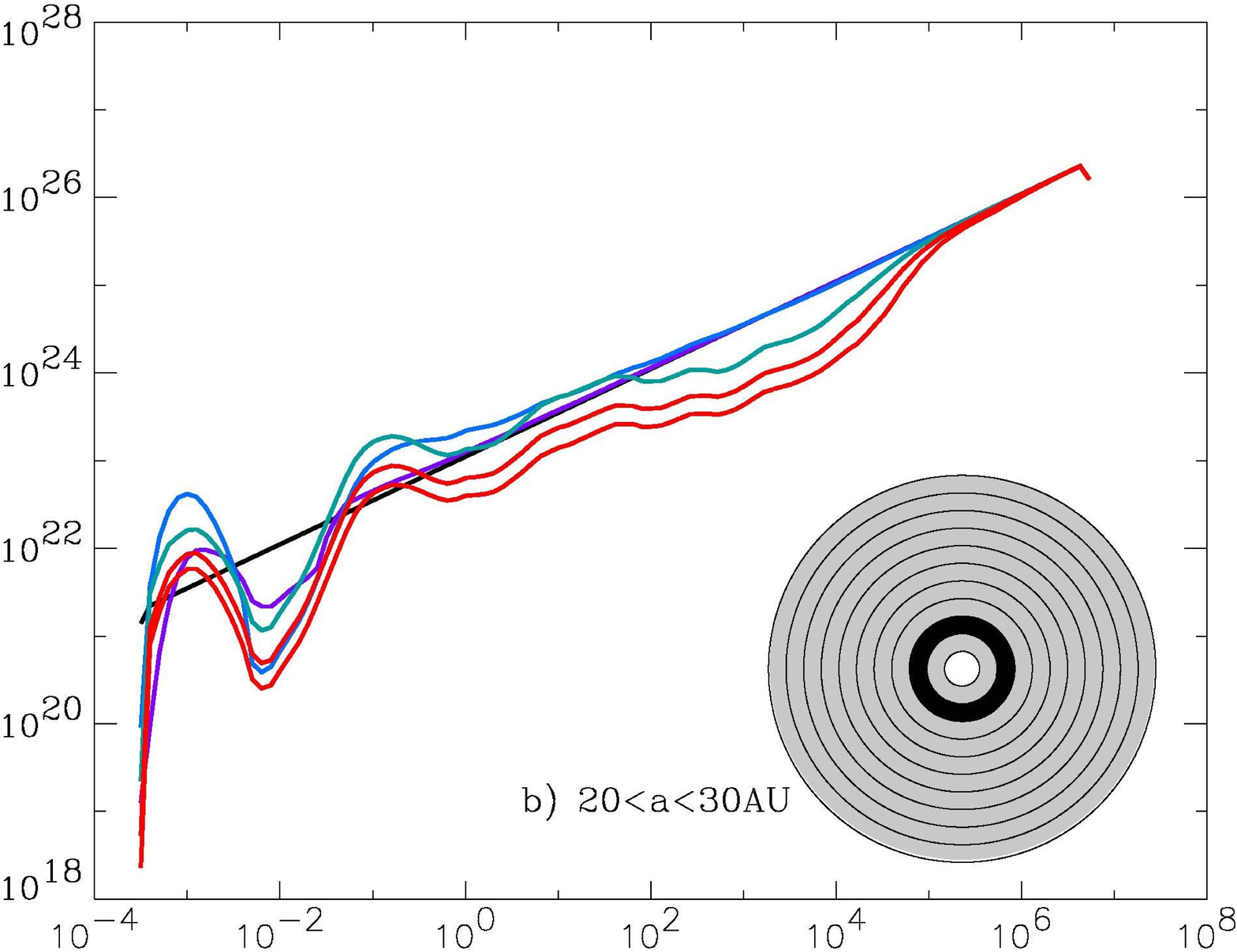}
}

\makebox[\textwidth]{
\includegraphics[width=\columnwidth]{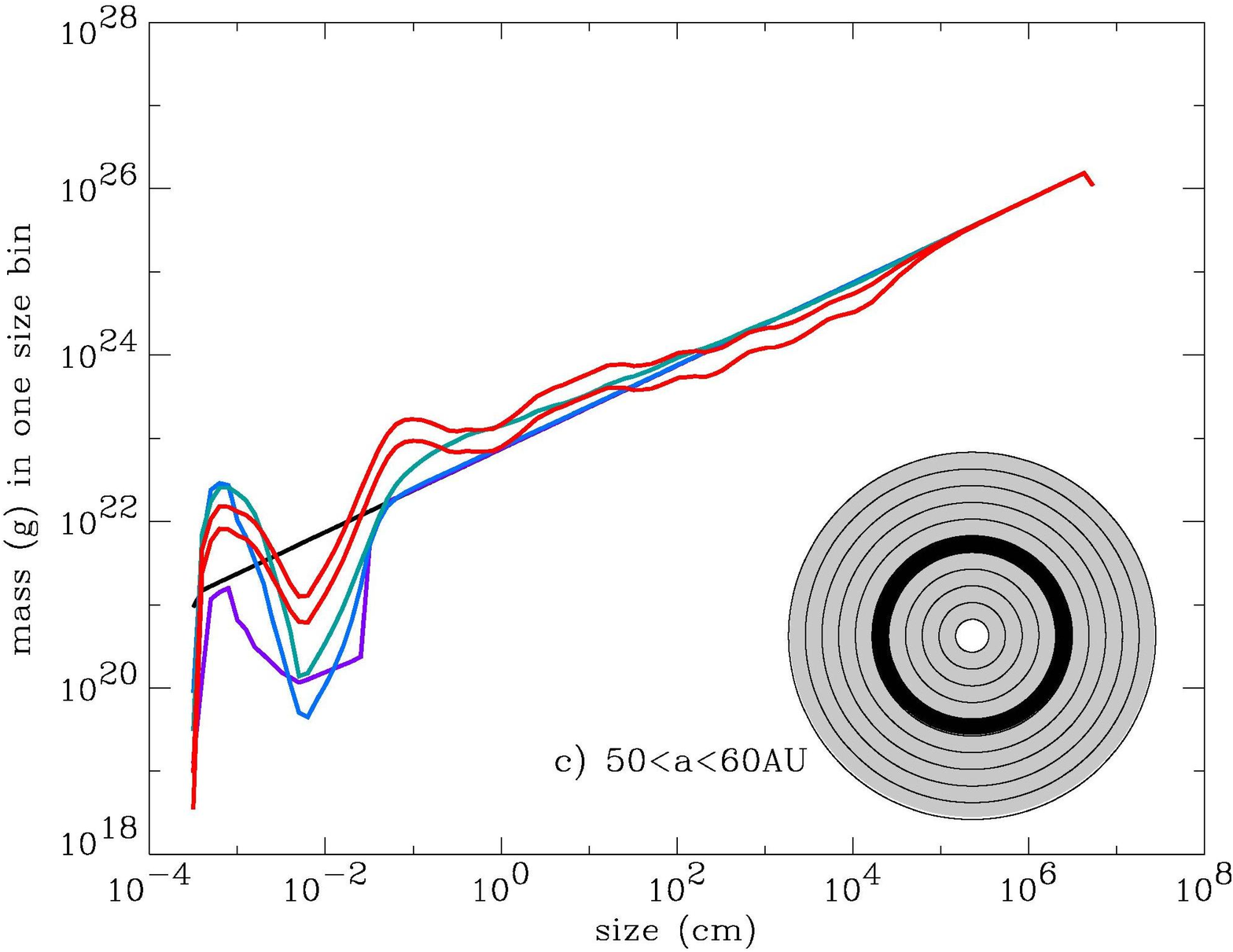}
\hfil
\includegraphics[width=\columnwidth]{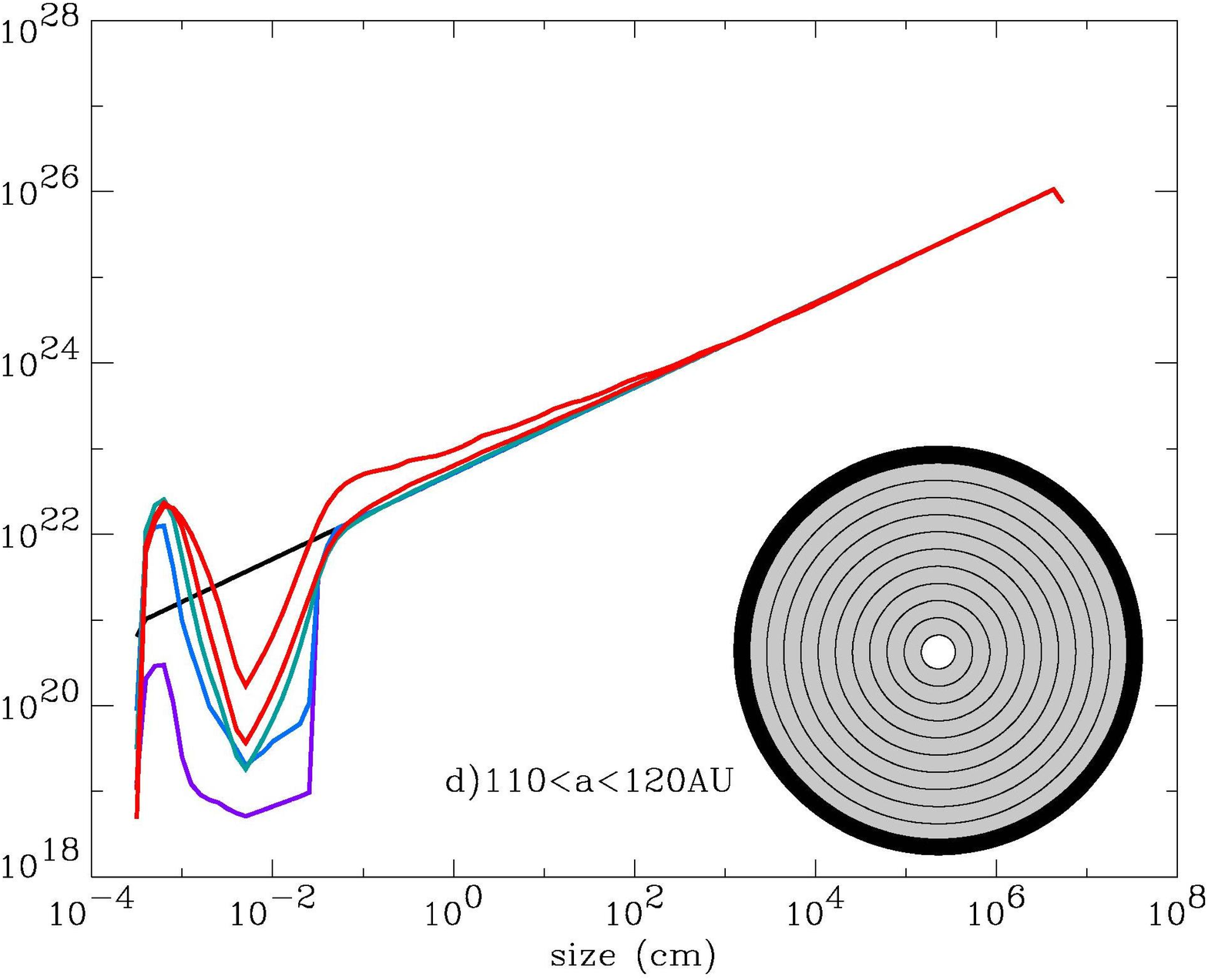}
}
\caption[]{Evolution of the size distribution for the
low--mass case ($M\dma{dust}=0.001 M_{\oplus}$).
}
\label{snapslmc}
\end{figure*}

The evolution of the size distribution for the low-mass case is displayed
in Figure~\ref{snapslmc}. The main difference with the $M\dma{dust}=0.1\,M_{\oplus}$
case is that the global evolution of the system is much slower. The slowing
down is logically of the order of the discs mass ratio (i.e. a factor of about
$100$). In the $a \la 70$\,AU region, after $10^{7}\,$years, the system
has reached a quasi steady--state relatively similar 
to the high mass case, with an overdensity of $\simeq 1.5\,R\dma{PR}$ grains,
followed by a depletion of sub--mm grains of approximately one
order of magnitude compared to the initial size distribution.

In the outermost regions, however, we observe a much deeper depletion
of sub--mm grains than in the high--mass case. This is because
the collisional--equilibrium, contrary to the inner disc regions,
has not been reached at $t=10^7$\,years:
the erosion of sub--mm grains, by high--$\beta$
particles coming from the inner regions, has already reached full efficiency
(after less than $10^{4}\,$yrs), while the production of new sub--mm
grains by erosion of larger objects occurs on much longer time--scales,
exceeding $10^{7}\,$yrs in the outer regions (see more detailed
discussion in the next section). This is clearly illustrated in
Fig.\,\ref{snapslmc}d, which shows that, in the outermost annulus,
the population of $>1$\,cm objects remains largely unaffected by collisional
processes after $10^7$\,years. 
In order to get a better idea (and despite of the huge CPU-time cost),
we decided to let this low-mass disc collisionally evolve for another
$t=10^{8}$\,years. As can be seen in Fig.\ref{snapslmc}d, at this later time
the quasi-steady state is almost reached in the outermost annulus, but
the second ``knee'' in the size distribution at $R\simeq 0.1$km is still not
visible yet. The full steady-state is here probably reached on timescales
of the order of $\sim 1$\,Gyr, which are presently out of the reach of
our numerical code.

\subsection{Collisional lifetimes}  \label{sec:colllife}

\begin{figure*}
\centering
\hbox to \textwidth
{
\parbox{0.5\textwidth}{
\includegraphics[width=\columnwidth]{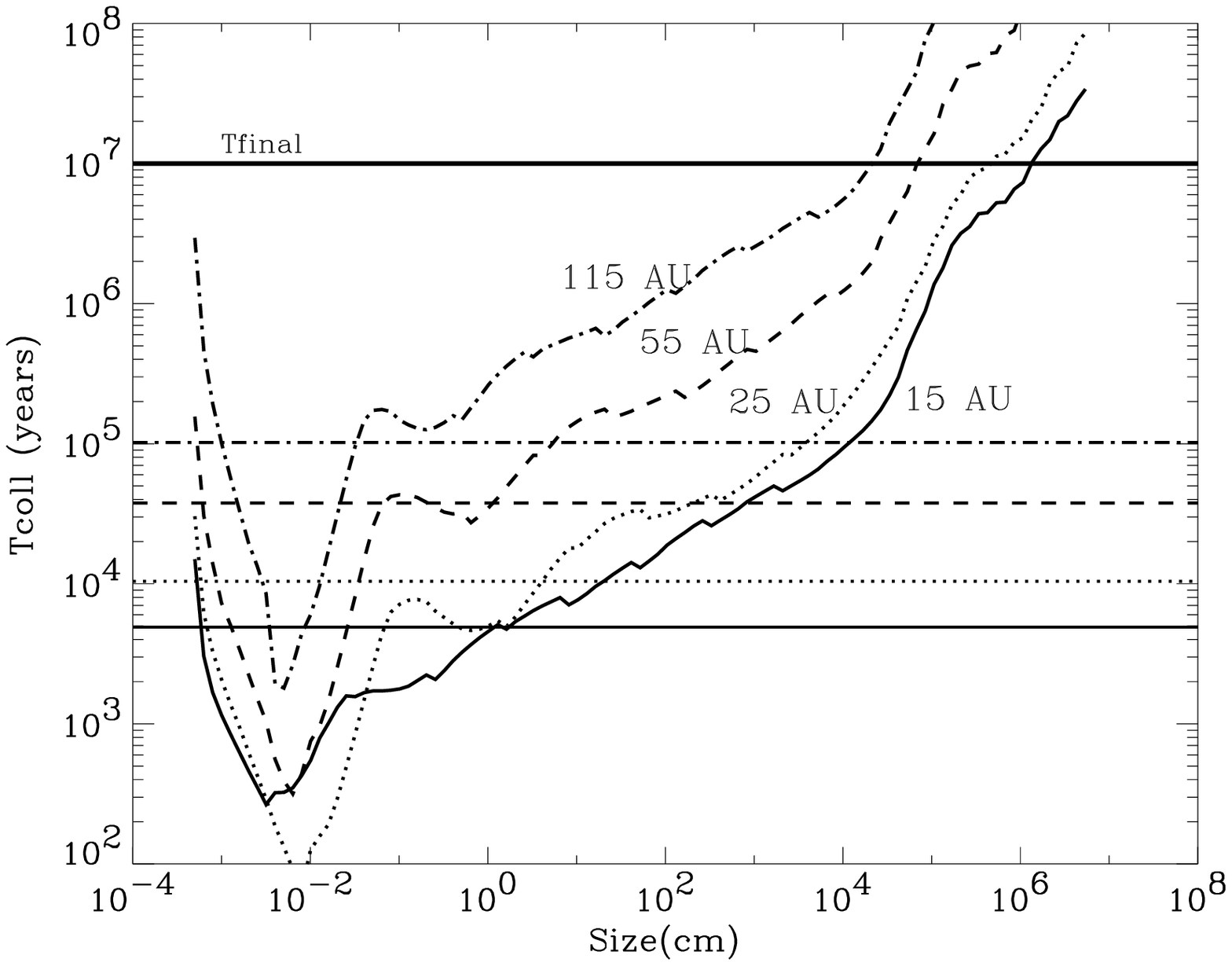}
}
\parbox{0.5\textwidth}{
\includegraphics[width=\columnwidth]{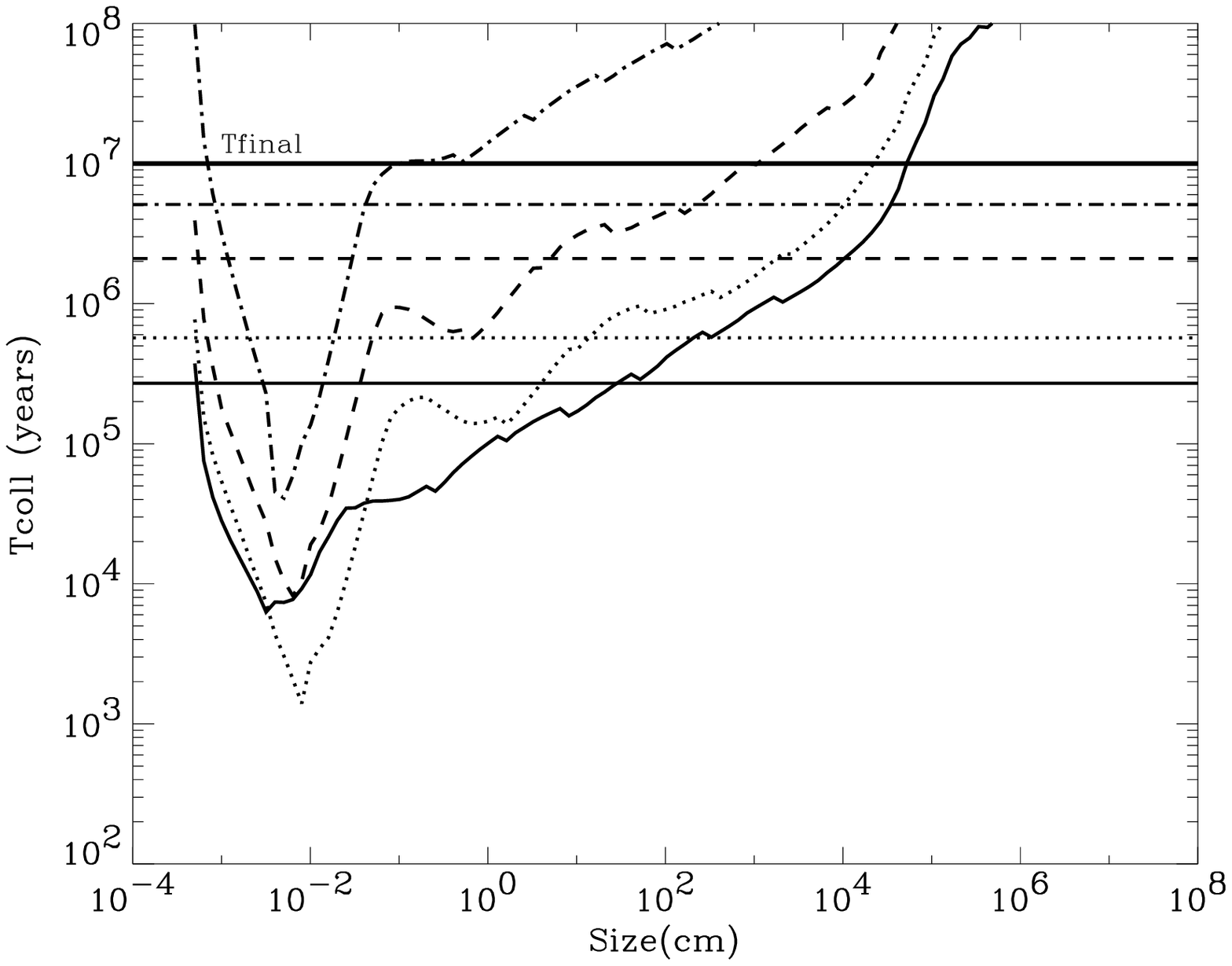}
}
}
\caption[]{Collisional lifetimes of the particles as a function of their size, at
$t=10^{7}\,$years, and at four different locations in the disc: $a=15\,$AU
(solid line), $a=25\,$AU (dotted line), $a=55\,$AU (dashed line), $a=115\,$AU
(dot-dashed line). The four horizontal lines are, for each distance, the collision timescales
deduced from the simplified $T^0\dma{coll}(a)=(\tau\Omega)^{-1}$ formula.
The $a$ values refer to the center of the annulus where the particle has been
produced. {\it Left panel}: high--mass case ($M\dma{dust}=0.1M_{\oplus}$),
{\it right panel}: low--mass case ($M\dma{dust}=0.001M_{\oplus}$)
}
\label{tcoll}
\end{figure*}

We define the collisional lifetime of a particle as the average time $t\dma{coll}$
it takes for the object to lose $100\%$ of its mass by
collisional processes. Let us point out that the collisional mass loss has
two origins: (i) catastrophic fragmentation, for which a particle loses by definition
$100\%$ of its mass at each fragmenting encounter, and (ii) cratering,
for which the particle is progressively eroded after each impact (excavated
mass $M\dma{cra}$ given in App.\,\ref{sec:appB3}).
The left and right panels of Fig.\,\ref{tcoll} display the values
of $t\dma{coll}(a,R)$,
after $10^{7}$\,years, at different radial locations in the disc, for both the
nominal high--mass and the low--mass cases.
Note that for particles placed on high--eccentricity
orbits by radiation pressure, $t\dma{coll}(a,R)$ is
the collisional lifetime of a particle initially produced at distance $a$,
when taking into account all the collisions this particle will suffer
in the different annuli it will cross on its eccentric orbit.

We discuss first the case of the high-mass disc (Fig.\,\ref{tcoll}, left panel).
For large objects, we obtain the predictable result that collision
lifetimes increase with increasing distances from the star. This results
from three concurring factors: particles number densities decrease
with $a$, dynamical timescales get longer, and impact velocities
lower, leading to less eroding impacts.
For objects in the dust size range, however, the situation is much more
complex, mainly because of the major influence the radial movements
of small high--$\beta$ grains have on the collisional evolution.
We find that the most short--lived particles are the ones
with $R\simeq 100\,\mu$m, which logically corresponds to the most
depleted population in the system (Fig.\,\ref{snapsh}).
For grains with sizes $R<100\,\mu$m, $t\dma{coll}$
very rapidly increases with decreasing sizes. The explanation for this
trend is twofold: first, destructive impactors in the $R<R\dma{PR}$ range
are strongly depleted, and second, many of these high--$\beta$
grains spend a large fraction of their eccentric orbits in
the empty region of the disc beyond 120\,AU.
This global trend of the $t\dma{coll}$ dependence with size could
to some extent be compared to the one obtained by \citet{stru06}
for the \mic\ system \citep[where stellar wind from the central M-type star
could play the same role as radiation pressure around A-type stars, see][]{aug06}.
These authors also found $\d t\dma{coll}/\d R>0$ for large grains
and a very sharp $\d t\dma{coll}/\d R<0$ gradient for smaller grains
(see Fig.1 of this paper). However, these similarities are
only qualitative (with major quantitative differences regarding
the turn--off size or the slope of the $\d t\dma{coll}/\d R$ laws) and 
should in any case be
taken with great care, since the \citet{stru06} estimates
were obtained  for a radially narrow system
and with a simplified analytical law for the collision rates and
outcomes.

The relative lifetimes between different regions of the disc follow
the logical trend in $\d t\dma{coll}/\d a>0$, with the important
exception of the innermost annulus, for which 
collisional lifetimes of small grains are relatively high, simply
because there is here no flux of destructive small (high--$\beta$)
impactors coming from further inside the system, contrary to the other annuli.
Note that the only objects having $t\dma{coll}>t\dma{final}=10^{7}\,$yrs are
the largest planetesimals, of size $R>0.2$\,km in the outermost regions,
and $R>3\,$km in the inner annuli. This means that, as a first
approximation, all sub--kilometre sized objects are collisionally 
evolved, i.e., no object other than the largest kilometre--sized
bodies are primordial. 

All these global trends are also valid for the low--mass case
(Fig.\,\ref{tcoll}, right panel). However, the fraction of primordial
objects is much higher than in the high--mass run. In the outer annulus,
for example, no object bigger than $\simeq1$\,cm has been
collisionally processed in $10^7$\,yrs
\footnote{For sake of comparison with the high-mass case, we
do here consider the same $t_{final}=10^{7}$years value for the low-mass
run, instead of the additional $10^{8}$years explored in Fig.3}.
This is in agreement with what was
pointed--out in the previous section, namely that the collisional
cascade did not fully develop in the outermost
regions after $10$\,Myr of evolution. For the dust--size range, however,
the result that all particles are collisionally processed during the
system's lifetime still holds. This is why the shape of the size
distribution is relatively similar in the high and low--mass runs.

As an interesting comparison, we also plotted on the graphs the collisional timescale
obtained by the formula $t^0\dma{coll}=(\tau\Omega)^{-1}$, where $\tau$ is the 
geometrical vertical
optical depth and $\Omega$ the angular velocity. This simplified
relation is indeed often used in the literature as giving an approximate estimate
of the collision lifetimes of the smallest grains.
As can be clearly seen, it proves to be a very poor match to our numerically
derived lifetimes. Differences can reach up to 2 orders of magnitudes in
the crucial $R<0.1$mm range.

\section{Parameter dependence exploration}  \label{sec:param}

\subsection{Dynamical excitation:  \,\,$\langle e\rangle$  } \label{sec:excite}
\begin{figure}
\includegraphics[angle=0,origin=br,width=\columnwidth]{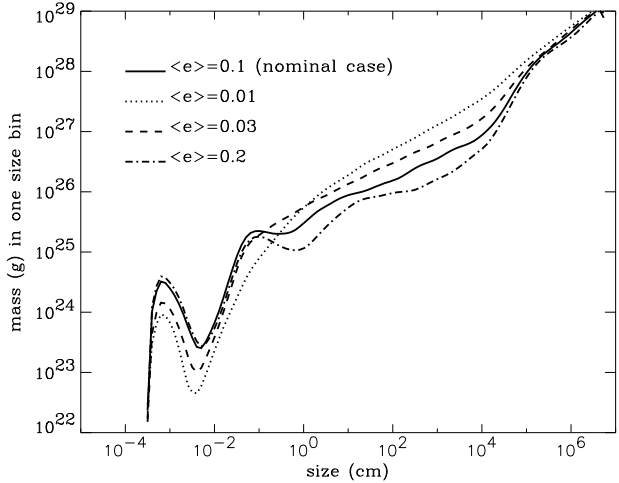}
\caption[]{
Impact of the disc's dynamical excitation on the size distribution after
$10^{7}\,$years of evolution. The figure shows the size distribution
of the {\it whole} system (i.e. all 11 annuli) assuming an initial
dust disc mass of $M\dma{dust}=0.1\,M_{\oplus}$ (high-mass case).
}
\label{excomp}
\end{figure}

The exact orbital distribution of particles in debris discs
is in general very poorly constrained. The only observational constraint
comes from measuring the disc's vertical thickness and deriving estimates
of orbital inclinations, but such constraints are scarce.
Edge-on discs represent the most favorable cases since $H/a$,
where $H$ denotes the vertical scale height, can be
directly measured. Five out of a dozen of spatially resolved
discs have this particular orientation: \bp, \mic,
\hdddd\ \citep{sch05}, \hdd\ \citep{kal06}, \hddd\ \citep{kal07}.
For the two most studied discs, only partial information is available.
\citet{kri05} find $H/a \la 0.04$ in the case of the \mic\ disc,
with ratios as small as $0.02$ close to the position
of maximum surface density. The \bp\ disc appears geometrically
thicker with $H/a$ ratios as large as $\simeq 0.1$ \citep{gol06}.
However, these measurements include the so-called disc warp which,
according to \citet{gol06}, might be due to a blend of
two separate, intrinsically thinner disc components inclined with
respect to each other by a few degrees. The \bp\ disc might then
in fact be less vertically extended than it appears to be.
The modeling and inversion of scattered light brightness profiles
of inclined, ring-shaped discs do not provide much more constraints.
The \hr\ and \hd\ rings for example, might have $H/a$ ratios as
large as about $0.1$ at the positions of maximum surface density,
but the actual ratios could be two times smaller \citep{aug99,sch06}.
As pointed out in section \ref{sec:init}, other estimates of the disc's vertical
thickness come from general theoretical arguments. Debris discs are indeed
believed to correspond to the late stages of planetary formation where
Lunar--to--Mars sized embryos dynamically excite the system. However,
this argument can only lead to rough order of magnitude estimates of
the dust's orbital elements.
It is thus important to explore different possible values of 
$\langle e\rangle$ and $\langle i\rangle$. Due to the
CPU--time consuming aspect of the simulations, we chose to restrict
ourselves to the high--mass system and perform
one additional ``dynamically colder'' case with
$\langle e\rangle = 0.03 = 2 \langle i\rangle$,
one ``very cold'' system with $\langle e\rangle = 0.01$
and one dynamically ``hotter'' case with
$\langle e\rangle = 0.2$.
A comparison between these three cases and the nominal $\langle e\rangle =0.1$
case is displayed in Fig.\,\ref{excomp}. For sake of clarity, we consider
here the whole system, summing up the contributions of all
radial annuli. 

Contrary to what could be intuitively expected, the depletion of
objects smaller than $1\,$mm is more pronounced in the
dynamically cold case $\langle e\rangle = 0.03$ (Fig.\,\ref{excomp}).
There are two concurring explanations for this apparent paradox.
On the one hand, the rate at which sub--millimetre grains are {\it eroded}
only weakly depends on the system's dynamical excitation. Indeed,
the velocity at which these grains are impacted by smaller micron--sized
particles is mainly imposed by the strong radiation force acting
on the latter and only weakly depends on the eccentricity of their parent
bodies' orbits.
On the other hand, the rate at which big grains are {\it produced}, by
impacts between larger objects, strongly depends on the system's
dynamical excitation, since these larger objects' orbits, and
thus their impact velocities, are insensitive to radiation pressure
effects.
As a consequence, the balance between production and erosion
of sub--mm grains is more negative for low
$\langle e\rangle$ values of the parent bodies orbits, hence
the more pronounced depletion.
For the ``very'' cold case, this effect is even more pronounced, and
one can witness a global general depletion of $all$ dust size grains,
while objects in the $\gtrsim 1$cm range are mostly unaffected
by any collisional evolution.

\begin{figure*}[ht]
\centering
\hbox to \textwidth
{
\parbox{0.5\textwidth}{
\includegraphics[width=\columnwidth]{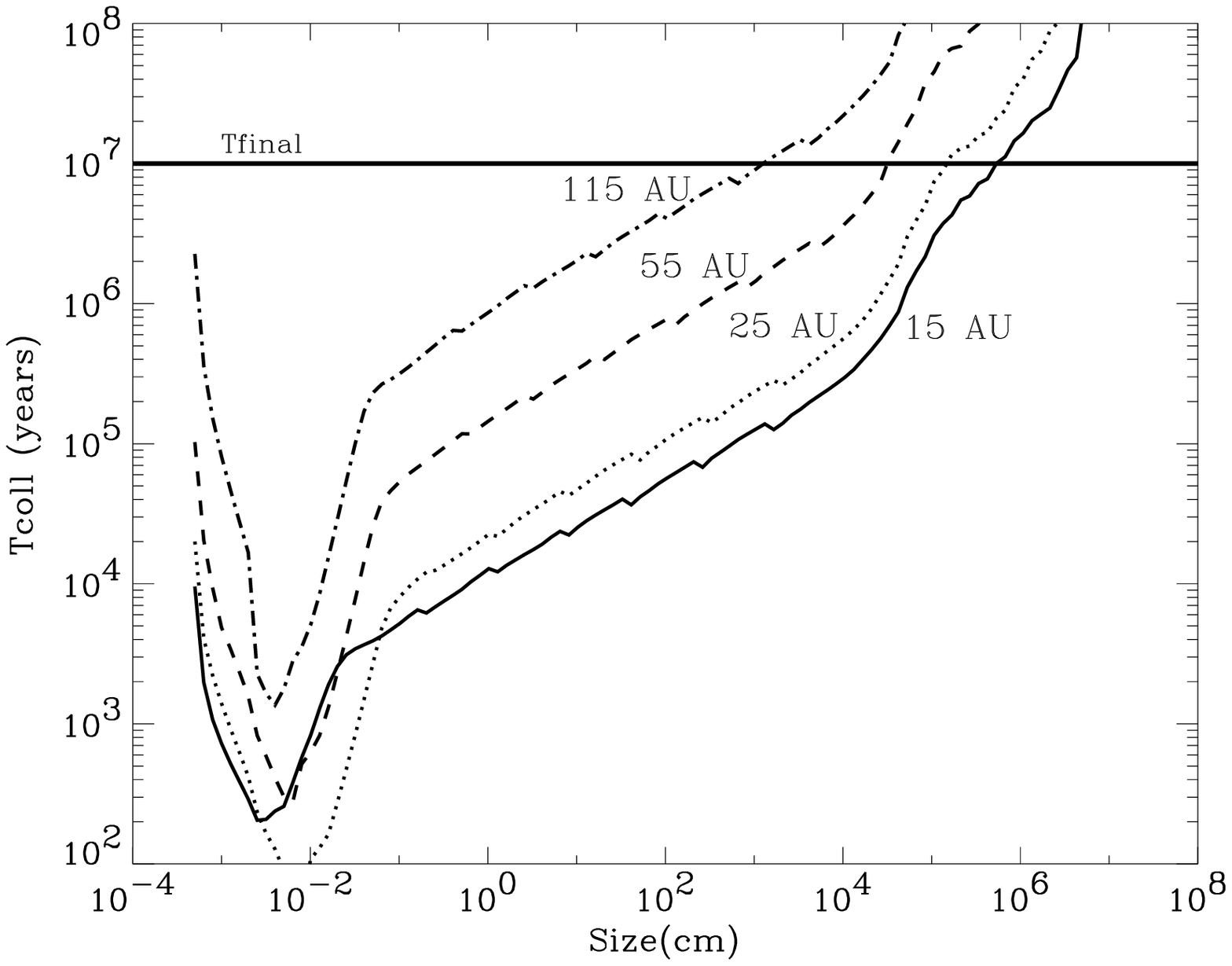}
}
\parbox{0.5\textwidth}{
\includegraphics[width=\columnwidth]{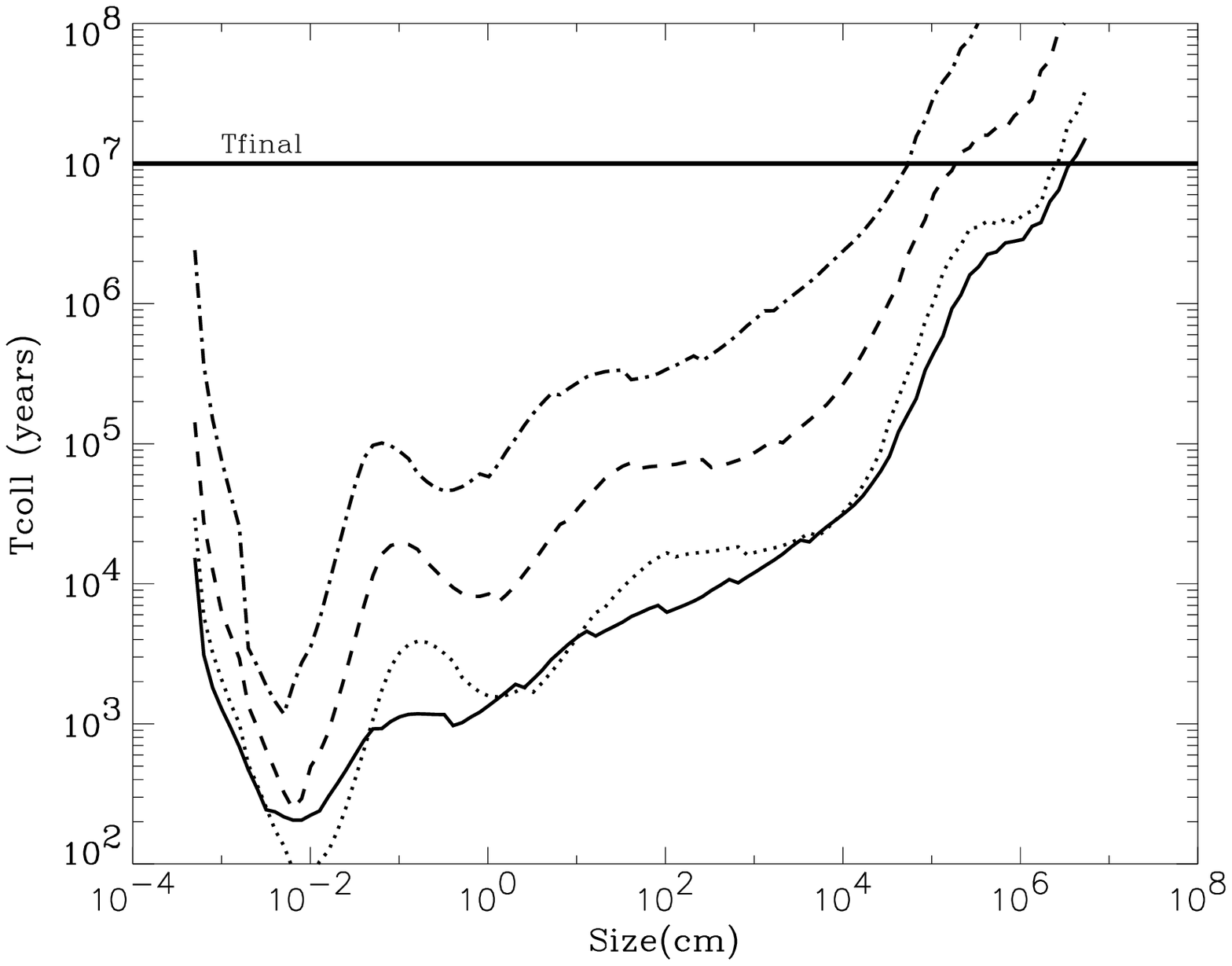}
}
}
\caption[]{Collisional lifetimes as
a function of particle size, at $t=10^{7}\,$years for
a dynamically ``cold'' case, $\langle e\rangle = 0.03$ ({\it left panel}),
and a dynamically ``hot'' case, $\langle e\rangle = 0.2$ ({\it right panel})
}
\label{tcolle}
\end{figure*}

For the high--excitation case, the depletion of sub--mm grains
is almost identical to the nominal case, which is here again
a direct consequence of the fact that the dynamics of the very
small grains is controlled by the radiation pressure force.
These results are clearly illustrated in Fig.\,\ref{tcolle}, showing
the collisional lifetimes in both the dynamically ``hot'' and ``cold''
cases. While $t\dma{coll}$ is roughtly  inversely
proportional to $\langle e\rangle$ for large ($\ga1$cm) particles,
the collisional lifetimes of small grains only weakly vary
with the average dynamical excitation in the disc.

\subsection{Mass of the star and $R\dma{PR}$ value}
\label{sec:radpre}

The nominal case considered in our simulations is that of a
$\beta$--Pictoris like star of mass $M_{*}=1.7\,M_{\odot}$ and
a corresponding radiation pressure cut--off size $R\dma{PR}=5\,\mu$m.
We explore here the $M_{*}$ and $R\dma{PR}$ parameters by considering,
in addition to the nominal case, two limiting cases: one G--star of
mass $1.1\,M_{\odot}$ with $R\dma{PR}=1\,\mu$m, i.e., the
lowest star mass for which compact silicate grains can reach
the $\beta = 0.5$ limit, and one Vega--like A0V star of
mass $2.5\,M_{\odot}$ and $R\dma{PR}=10\,\mu$m. All $R\dma{PR}$ values
have been derived using the \citet{grig07} algorithm.

\begin{figure}
\includegraphics[angle=0,origin=br,width=\columnwidth]{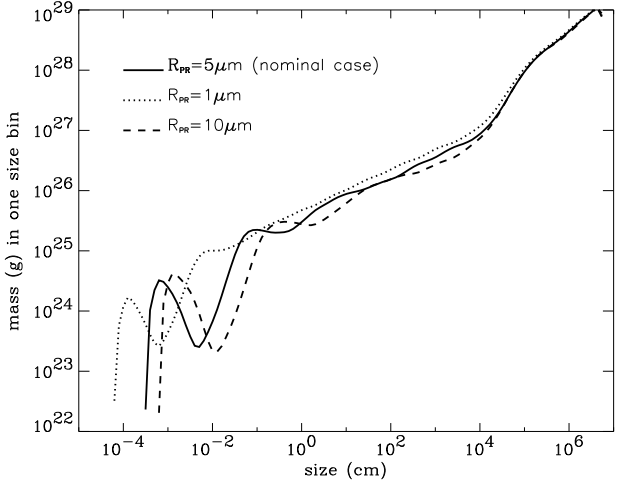}
\caption[]{
Impact of the radiation pressure cut--off size value $R\dma{PR}$
on the size distribution of the {\it whole} system (i.e. all 11 annuli)
at $t=10^7\,$years for the high--mass case ($M\dma{dust}=0.1\,M_{\oplus}$).
{\it Solid line}: nomical case, $M_{*}=1.7\,M_{\odot}$ and $R\dma{PR}=5\,\mu$m,
{\it dotted line}:$M_{*}=1.1\,M_{\odot}$ and $R\dma{PR}=1\,\mu$m,
{\it dashed line}: $M_{*}=2.5\,M_{\odot}$ and $R\dma{PR}=10\,\mu$m
 }
\label{miccomp}
\end{figure}

As appears clearly on Fig.\,\ref{miccomp}, the size distributions
for all three systems in the ``dust'' grains size range ($R<1\,$cm) are
relatively similar. The profiles are shifted in size with respect
to each other, reflecting the difference in $R\dma{PR}$ values.
Interestingly, the location of the overdensity of smallest grains
is always given by the relation $R\simeq 1.5\,R\dma{PR}$, while the most
pronounced depletion is always obtained for $R\simeq 10-50\,R\dma{PR}$.
However, the amplitude of this depletion increases with increasing
$R\dma{PR}$ values (i.e. star masses). This is because, in the strength
regime, smaller grains are more resistant to impacts than bigger ones,
which implies that an impact between, say, a $R=1.5\,R\dma{PR}$ dust grain
and a $R=100\,R\dma{PR}$ object is more erosive for larger values
of $R\dma{PR}$. Furthermore, for a more massive star, impact velocities
are higher (for the same orbital parameters), which also leads to more destructive
collisions.

\subsection{Initial density profile}
\label{sec:initdens}

\begin{figure}
\includegraphics[angle=0,origin=br,width=\columnwidth]{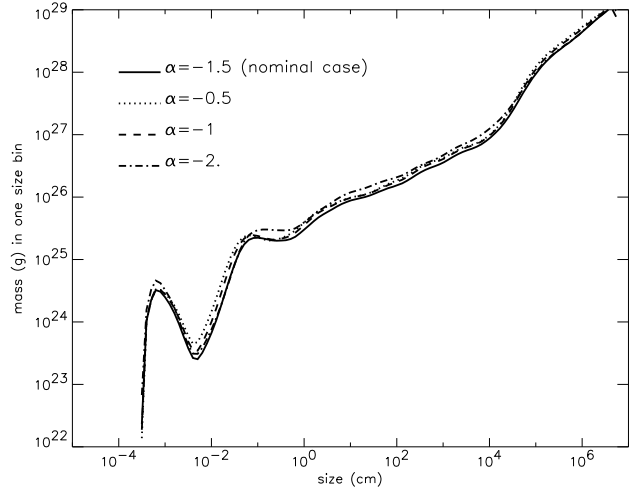}
\caption[]{Size distribution, at $t=10^{7}\,$years, for the $whole$ system
(all 11 annuli), for 4 different initial surface density distributions
$\Sigma(a) \propto a^{\alpha}$, keeping the system's total mass constant.
}
\label{dissize}
\end{figure}

We have considered as a standard case a system following a standard
MMSN spatial distribution in $\Sigma(a) \propto a^{-1.5}$.
However, in order to check the robustness of our results,
other indexes for the $\Sigma(a)\propto a^{\alpha}$ dependence have been
explored. 
Fig.\,\ref{dissize} shows that the global size distributions
within the system only weakly depends on the initial $\Sigma(a)$ power law.
The only noticeable trend is a slight damping of the wavy
distribution in the $R\la$1cm range for flatter $\Sigma(a)$ profiles.
This result is logical since we have seen in section \ref{sec:nomin}
that, in a given region of the disc, the evolution of the sub--mm
grains is mainly imposed by the flux of high--$\beta$ particles
coming at high radial velocities from the inner regions.
The influence of these inner--disc born grains should logically
diminish for less steep $\Sigma(a)$ profiles, for which their relative
abundance compared to the local population is smaller.
However, these differences between the different $\Sigma(a)$ cases
remain limited in amplitude and all size distributions remain
very close to the result of the nominal case.

\subsection{Collision outcome prescription}
\label{sec:collpres}

\begin{figure}
\includegraphics[angle=0,origin=br,width=\columnwidth]{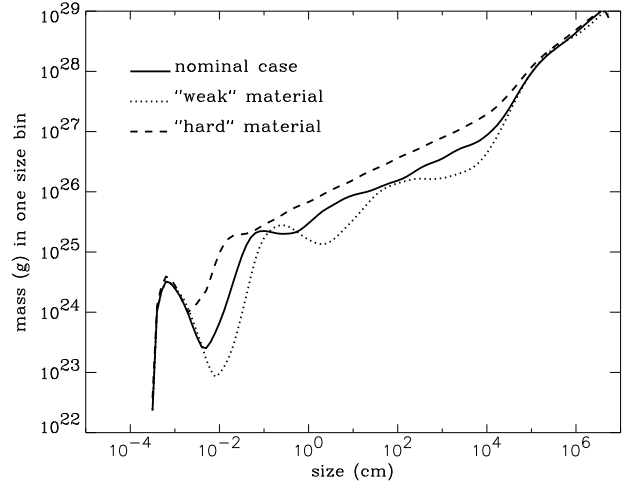}
\caption[]{Impact of the outcome collision prescription on the size distribution
at $t=10^{7}\,$years and for the {\it whole} system (i.e. all 11
annuli). Three different collision-outcomes prescriptions have been
assumed: nominal case (solid line), ``hard material'' case (dashed line) and
``weak material'' case (dotted line). See text for details.
 }
\label{compcoll}
\end{figure}

As discussed at length in the Appendix\,\ref{sec:appB}, the collision outcome
prescription is a poorly constrained parameter, first because of uncertainties
regarding the chemical composition and structure of the grains and planetesimals
in debris discs, and second because of significant differences between the predictions
of all existing models. Our nominal case assumes a sublimation distance for ices
$a\dma{sub}=20\,$AU, the
\citet{benz99} prescription for the critical specific energy $Q^{*}$ for
silicates and $Q^{*}\dma{ice}=Q^{*}\dma{sil}/5$, and the \citet{kos01} formula
for crater--excavated masses $M\dma{cra}$ for ices and silicates (see
Appendix\,\ref{sec:appB}). In order to explore how our results depend on
the collision prescription, we have performed the two following additional runs:
\begin{itemize}
\item
one ``hard'' material run, assuming the
\citet{benz99} and \citet{kos01} prescriptions for
compact silicates hold for the entire disc (i.e. numerically
setting $a\dma{sub}=\infty\,$AU)
\item
One ``weak'' material run, where we assume the $Q^{*}$
prescription of \citet{kriv06}, and a value of $M\dma{cra}$ five times
higher than in the nominal case.
\end{itemize}
The results are displayed in Fig.\,\ref{compcoll}

As could be logically expected, the wave--like
structure is much less pronounced for the ``hard'' material run. As a matter
of fact, only the first wavy feature, affecting the smallest grains,
is clearly visible, and its amplitude is damped by a
factor $\simeq 3$ compared to the nominal case. Moreover, the size
for which the strongest depletion is reached is shifted from
$R\simeq 20R_{PR}=100\,\mu$m to $R\simeq 4R_{PR}= 20\,\mu$m. 
For the ``weak'' material run, the exact opposite is observed: pronounced
wavy--features propagate up to the largest sizes, and the amplitude of the
depletion of sub--mm grains is significantly increased and reaches almost
two orders of magnitude. Contrary to the hard--material run, the
depletion is now shifted towards bigger grains as compared to the nominal
run. The weak--material run partially 
resembles the results of \citet{kriv06}, which
is logical considering that we took identical $Q^{*}$ values, but
differences are observed, which can probably be attributed to the fact
that cratering impacts are here taken into account.

A comparison between Fig.\,\ref{compcoll} and all other parameter exploration
runs of Figs.\,\ref{excomp} to \ref{dissize} clearly shows 
that the collision--outcome prescription is the most crucial
parameter the final size--distribution depends on.
Unfortunately, this parameter is probably
the most poorly constrained in the present problem.
As described at length in the Appendix, particular attention
has been paid here to this crucial issue. We have tried to
improve on most previous studies (including TAB03) 
and consider an upgraded model incorporating the most relevant
available data for the $Q^{*}$ as well as fragmentation and, more
specifically, cratering prescriptions. Nevertheless, 
large uncertainties remain.
Firstly, important grain properties, which are crucial
for understanding their response to impacts (ice fraction, porosity, etc...),
remain poorly constrained for most debris discs.
Secondly, even if all grain
characteristics were fully known, it remains to see to which
extent collision outcome energy--scaling models
(even the more advanced version considered here), mostly obtained by
experiments on cm--to--decimetre sized targets, might apply over
such a wide size range, especially for very small micron--sized grains.
There is to our knowledge no fully reliable data
on what the outcome of a collision between, say, a 5$\mu$m grain and a
0.1mm target at 500m.s$^{-1}$ ``really'' is.
Basically, it all comes down to how soft or hard (with
respect to a collisional event) particles
in the $<1$cm range are, and how these characteristics
might vary with size. In this respect,
we believe our nominal case collision prescription to be
the most reliable one given the (still limited) current knowledge on this
complex problem. Nevertheless, significantly different collisional
behaviours cannot be ruled out. Fig.\,\ref{compcoll} probably
gives a good idea of realistic boundaries for 
the limiting ``hardest'' and ``weakest'' material cases,
showing that the waviness of the size distribution
decreases with increasing collisional resistance of the objects.

\section{Spatial distribution and dust to planetesimals mass ratios}  \label{sec:spat}

\subsection{Radial distribution}

\begin{figure}
\includegraphics[angle=0,origin=br,width=\columnwidth]{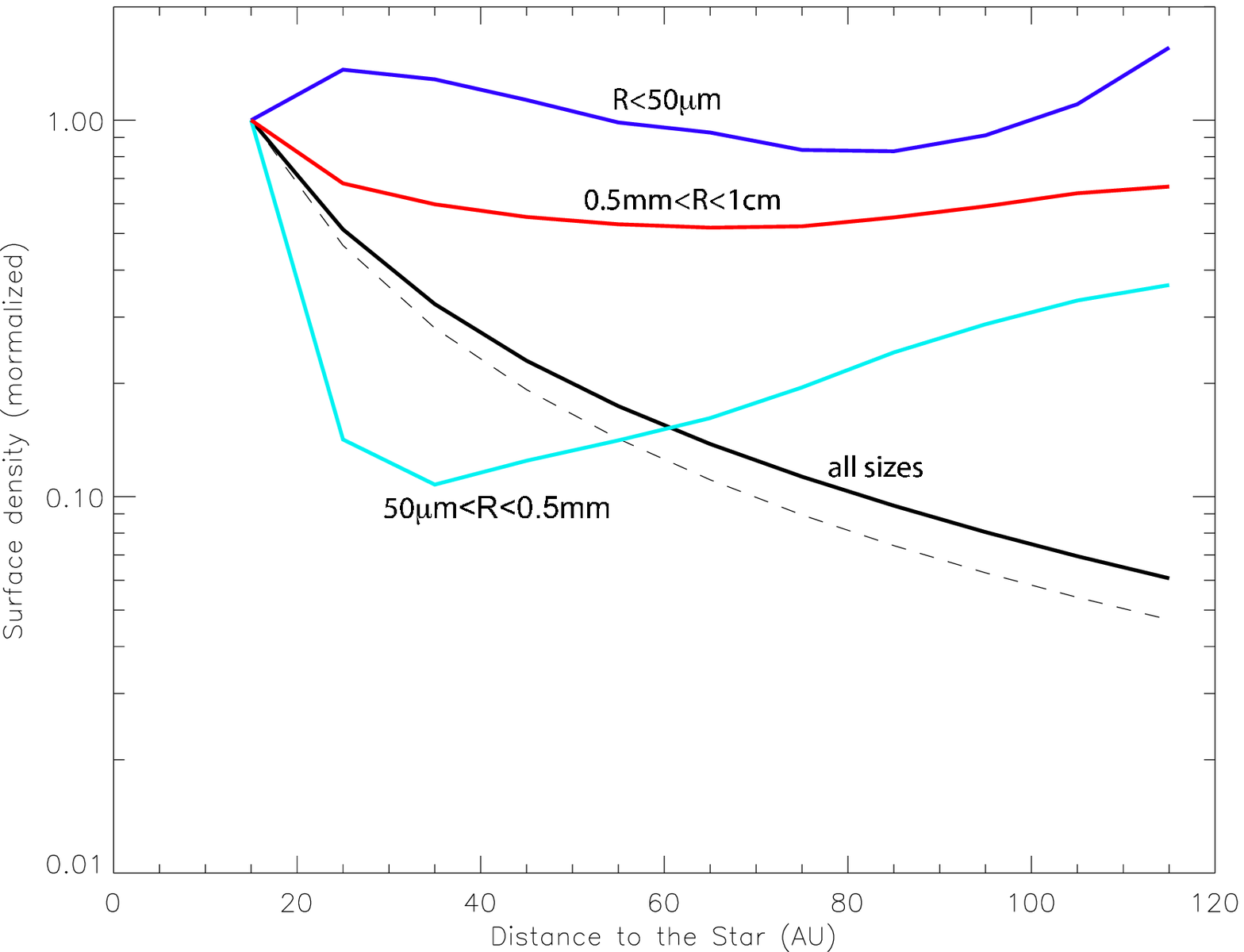}
\caption[]{High--mass disc. Radial distribution, at $t=10^{7}\,$years,
of the mass surface density for different object sizes.
For each size range, all surface densities are renormalized to the surface density
in the first annulus.
The dashed line
represents the theoretical distribution should
a MMSN power law in $a^{-1.5}$ hold starting at the innermost annulus.
 }
\label{surfcomp}
\end{figure}
\begin{figure}
\includegraphics[angle=0,origin=br,width=\columnwidth]{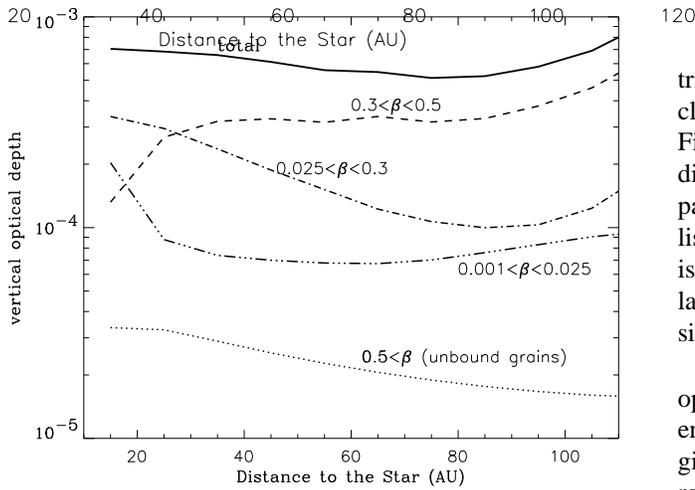}
\caption[]{High--mass disc. Radial distribution, at $t=10^{7}\,$years,
of the geometrical vertical optical depth for different size ranges, parametrized
by their $\beta$ parameter.
 }
\label{opt}
\end{figure}

For sake of clarity, we consider here only the nominal high--mass run.
Fig.\ref{surfcomp} clearly shows that the spatial distribution
significantly departs from the MMSN profile for all objects in
the ``dust'' size range ($<1$cm).
As could be logically expected, the strongest departure from
the initial MMSN profile is obtained for grains in the sub-mm size
range. For this population, the sharpest feature is a density drop in the regions
just outside the first annulus. 
This drop is easily understandable and is due to the inter-annuli
interactions already described in 3.1.1:
in the innermost annulus, only $locally$ produced small grains can
erode sub--mm particles, but such 
locally produced small grains, blown out by
radiation pressure on unbound or very elliptical orbits, have not the time
to be accelerated to high velocities, which limits their destructive or
erosive power. In all other annuli, on the contrary,
small grains coming from the inner regions impact local
bigger grains at very high velocities and are able to deplete them
more significantly.

For small grains in the $<50\,\mu$m range,
the radial distribution is very flat,  even flatter than the one which should be
expected in a steady flow of outgoing unbound particles, where
simple mass conservation considerations lead to $\Sigma(a) \propto a^{-1}$
\citep[e.g.][]{su05}. This profile cannot be explained by
simple blow out of unbound particles since most
of the grains in the $<50\,\mu$m range are on $bound$ orbits ($R\dma{PR}=5\,\mu$m
for our nominal case).
On the other hand, the mass surface density distribution of the total system
(all particle sizes) is still relatively close to a classical MMSN
profile in $a^{-1.5}$ (solid black line in Fig.\,\ref{surfcomp}). This is not
an unexpected result, since the bulk of the disc's mass is still
contained in the biggest, kilometre--sized particles, which are only
marginally affected by specific collisional behaviour
of the smallest grains.
Therefore, there exists a major discrepancy between the spatial distribution
of the largest  undetectable objects and that of the grains in
the dust--size range, i.e. those accessible to observations.

Another interesting result concerns the geometrical vertical optical depth $\tau(a)$.
Fig.\,\ref{opt} shows the respective weight of different grain
populations. We see that, except for the innermost regions, $\tau(a)$
is completely dominated by grains from a very narrow size range 
of $\alpha$--meteoroids just above
the blow--out limit $\beta=0.5 \Leftrightarrow R=R\dma{PR}$.
Of course, even with a standard power law distribution in
$\d N\propto R^{-3.5}\d R$, the optical depth should be dominated
by small objects, since $\int \tau(R)\d R\propto
\left(R^{-0.5}_{2}-R^{-0.5}_{1}\right)$.
However, this tendency is much more pronounced here.
As a matter of fact, when averaged over the whole system, it
can be shown that 50\% of the total optical depth is due to bodies
in the $0.3<\beta<0.5\Leftrightarrow R\dma{PR}<R<1.6R\dma{PR}$ range.
For a Dohnanyi profile, the size range containing 50\% of
the total optical depth is much broader: $R\dma{PR}<R<4R\dma{PR}$.
\begin{figure}
\includegraphics[angle=0,origin=br,width=\columnwidth]{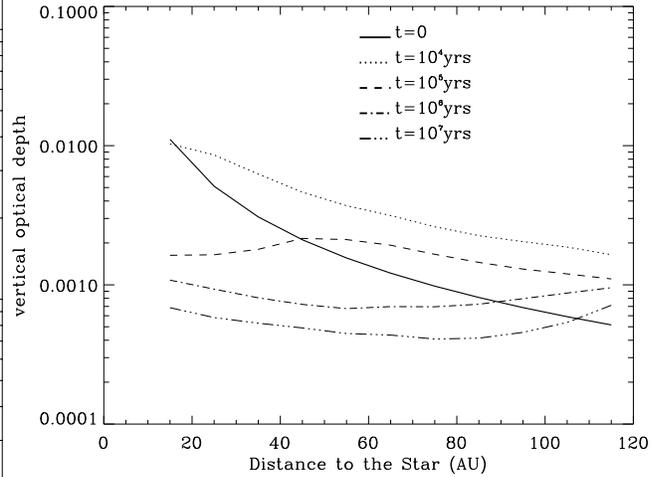}
\caption[]{High--mass disc. Evolution of the vertical optical depth profile
with time.
 }
\label{optev}
\end{figure}
The temporal evolution of the $\tau(a)$ profile is also
of interest. As Fig.\,\ref{optev} clearly shows, it rapidly
settles (in a few $10^{5}$yrs) to a relatively ``flat'' radial
profile, much flatter than the initial $a^{-1.5}$ one. This
flattening is due to several mutually connected factors. The main
one is due to what has been previously outlined, namely
that the optical depth is dominated by grains from a narrow size range just
above $R\dma{PR}$. These very small grains are very quickly placed
on very eccentric orbits, and will thus spend
most of their orbits \emph{outside} their annulus of production. 
As a consequence, small high-$\beta$ grains will naturally tend to
be depleted in the inner regions and pile-up in the outer ones.
In addition to this, the collisional erosion of bigger dust grains
in the $\sim 0.05\,$mm to $1\,$mm range, which make up most of
the mass ``reservoir'' from which smaller high-$\beta$ grains are
collisionnaly produced, is faster in the inner regions than in the outer ones
(see Fig.\,\ref{snapsh}). For the innermost annulus, this significant
mass erosion is even observed for the biggest planetesimals at the
upper end of the size distributions (which get depleted by a factor
$\sim 2$ in $10^7$yrs). It should be noted that the erosion of the $\sim 0.05\,$mm
to $1\,$mm grains is sensitive to the collisional prescription: neglecting for instance
cratering impacts leads to a much slower evolution of this population and
thus a much slower flattening of the profile.

\subsection{Link between dust and planetesimals}
\begin{table}
\begin{center}
\caption[]{Relative mass fraction $M_{\mu m}$ contained in the smallest
($R<20\,\mu$m) grains, $M_{mm}$ in the biggest dust particles
($0.1$\,mm$<R<1\,$cm), and $M_{big}$ in the
biggest $100\,$m$<R$ bodies, for all numerically tested cases and for
a standard $\d N\propto R^{-3.5}\d R$ size distribution.
All fractions are normalized to $M_{mm}$.}
\label{fracmass}
\begin{tabular}{lcc}
\hline
Run & $M_{\mu m}/M_{mm}$ & $M_{big}/M_{mm}$ \\
\hline
Nominal case & 0.0678 & 3770.8 \\
Low mass case & 0.0744 & 1937.5\\
$\langle e \rangle$=0.01 & 0.0318 &  8515.7\\
$\langle e \rangle$=0.03 & 0.0278  &3889.5\\
$\langle e \rangle$=0.2 & 0.1136  & 4476.2\\
Hard material case & 0.0399 & 2137.7\\
Weak material case & 0.0840 & 4112.3\\
$\Sigma(a)\propto a^{-0.5}$& 0.0639 & 3655.3\\
$\Sigma(a)\propto a^{-1}$& 0.0652   & 3634.2\\
$\Sigma(a)\propto a^{-2}$&0.0705  & 2958.1\\
\hline
$\d N\propto R^{-3.5}\d R$ distribution& 0.0675 & 3510.8\\
\hline
\end{tabular}
\end{center}
\end{table}

As described at length in the introduction,
an important issue is the link between the observed dust population
and the unseen bigger parent bodies. 
We report in Table\,\ref{fracmass} the respective masses of 3
representative populations: 
\begin{itemize}
\item
the smallest $R<20\,\mu$m grains,
i.e., the population containing most of the optical depth
\item
all grains in the $0.1\,$mm to $1$\,cm range, i.e., where most of the observable
``dust'' mass is
\item the biggest objects in the $R>100$\,m range
\end{itemize}
Surprisingly enough, the respective masses between these 3 populations
never drastically differ from their values in a standard Dohnanyi
distribution. Both $M_{\mu m}/M_{mm}$ and $M_{big}/M_{mm}$
stay within a factor$\sim$3 above or below the reference values
derived by integrating a $\d N\propto R^{-3.5}\d R$ power law.
As a consequence, despite the strong wavy features of the size distributions,
the link between the $total$ amount of observed dust and unseen bigger bodies
can be, as a first approximation, derived using a simple Dohnanyi
power law.

\section{Impact on the observations} \label{sec:obs}

\subsection{Scattered light surface brightness profiles} \label{sec:profs}

\begin{figure*}[ht]
\centering
\hbox to \textwidth
{
\parbox{0.5\textwidth}{
\includegraphics[width=\columnwidth]{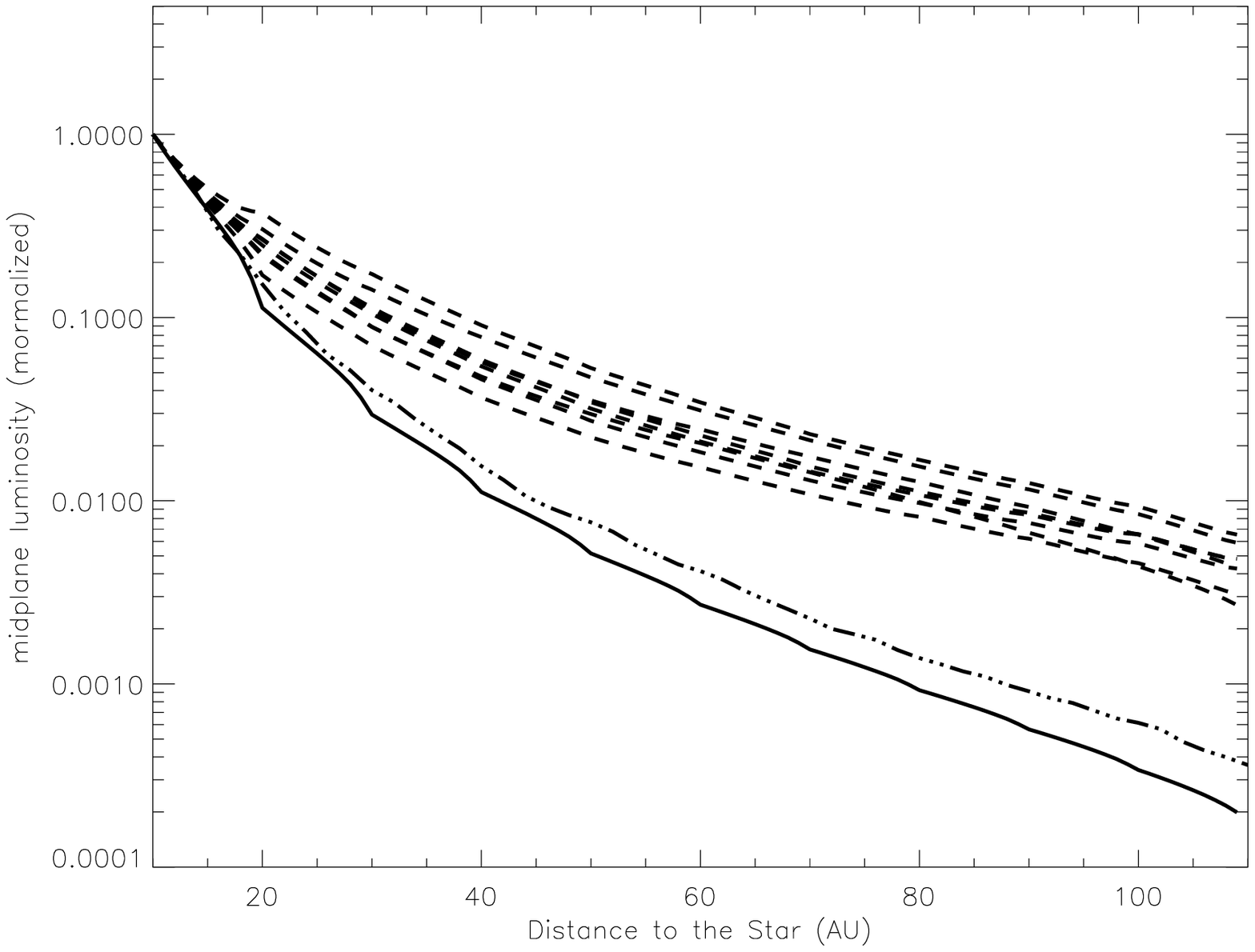}
}
\parbox{0.5\textwidth}{
\includegraphics[width=\columnwidth]{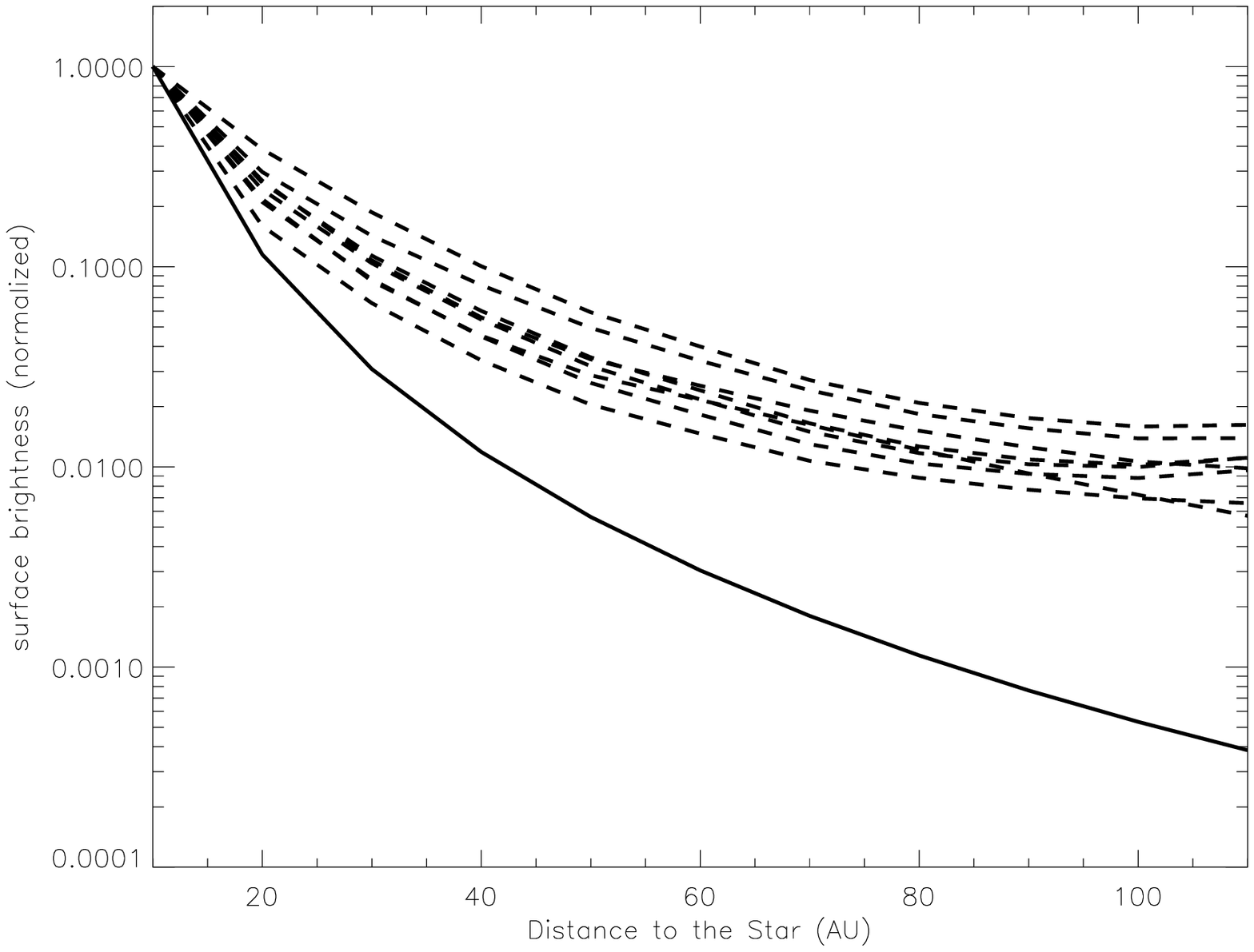}
}
}
\caption[]{
Synthetic luminosity profiles in scattered light
(isotropic scattering), for the final states
of all cases explored in section \ref{sec:param} (dashed lines); with the 
exception of those with alternative initial density profiles.
All profiles have been renormalized to 1 at 10AU.
The full line represents the theoretical brightness profile should 
a MMSN power law in $a^{-1.5}$ hold, for all grain sizes,
starting at the innermost annulus.
The dashed--dotted line represents the theoretical profile obtained when
assuming a MMSN surface density in $a^{-1.5}$ for all bigger grains on Keplerian
orbits, and assuming that all smaller radiation--pressure affected grains,
up to $\beta < 0.5$, are produced by the bigger grains
following a size distribution in $R^{-3.5}$ (see text for details).
{\it Left panel}: disc seen edge—on, i.e., radial mid-plane profiles
{\it Right panel}: disc seen head-on, i.e., average surface brightness.
 }
\label{profrad}
\end{figure*}

\subsubsection{Nominal case}

We consider here the two limiting cases of edge--on and head--on viewed systems.
For sake of simplicity, we have assumed gray scattering and we display results only
for the pure isotropic scattering case. However, other scattering phase
functions have been explored, and we verify that the results presented
hereafter, in particular regarding the departure from the initial profiles,
still hold for all explored cases. We furthermore assume the disc vertical
scale height $H$ varies linearly with the distance to the star.

For the edge--on viewing case,
Fig.\,\ref{profrad} (left panel) shows that, surprisingly,
the final mid--plane surface brightness (hereafter $S\!B$) profiles only
weakly vary with the parameters explored in the different runs. 
For all 9 cases considered, the scattered light radial profiles
approximately follow a power law in 
$S\!B\dma{edge}(a) \propto a^{b}$
with $-2.4<b<-2.1$.
This is very different from what is obtained for a theoretical system
where all bodies follow the initial $\Sigma(a) \propto a^{\alpha}$
radial distribution and a Dohnanyi--like distribution holds over
the whole size range, for which we get $b_0\simeq-3.4$
(for $\alpha=-1.5$), close to the theoretical value of $-3.5$ \citep[e.g.][]{nak90}. 
We shall from now on refer to this theoretical disc, which in fact corresponds
to the situation at $t=0$ in our simulations, as the ``static'' case, with
$S\!B\dma{edge}(a) \propto a^{b_0}$ and $b_0 = \alpha - 2$ (again assuming $H\propto a$).
A similar result holds for the head--on case, for which average $S\!B$ profiles
also strongly depart from the MMSN case (Fig.\,\ref{profrad}, right panel).
In other words, $S\!B$ profiles cannot be simply derived
by assuming the simplest hypothesis that dust grains follow the same
spatial distribution as larger parent bodies (for which the
initial $\Sigma(a)\propto a ^{\alpha}$ profile still holds).

Interestingly, neither can these $S\!B$ profiles be
derived by assuming the seemingly more advanced hypothesis that all
small (i.e. radiation pressure affected) particles have eccentric
orbits with their
periastron coinciding with the big particles distribution
and their number density being derived by the classical
$\d N \propto R^{-3.5}\d R$ size distribution.
This possibility has been checked following the method of \citet{aug01}
and \citet{theb05}:
we run a simple deterministic orbital integration where
$0<\beta<0.5$ grains are randomly produced from an initial
parent body population (following here a surface density profile
in $a^{-1.5}$). The distributions of all grains of a given $\beta$
are then obtained by phase mixing of their orbits and
the total resulting surface density by weighting each contribution
according to a Dohnanyi size distribution.
The resulting mid--plane $S\!B$ profile is shown in
Fig.\,\ref{profrad} (triple dot-dashed line).
Although it is a slight improvement over the pure ``static'' case, it is
still far from all synthetic profiles obtained with our collisional evolution
code. This means that the profile flattening is not simply due to
the geometrical spread of high--$\beta$ grains on eccentric orbits. It
is the consequence of the more complex effects these movements of
radiation--pressure affected grains have on the collision production
and destruction rates of dust grains in the different regions of the disc.

\begin{figure}
\includegraphics[angle=0,origin=br,width=\columnwidth]{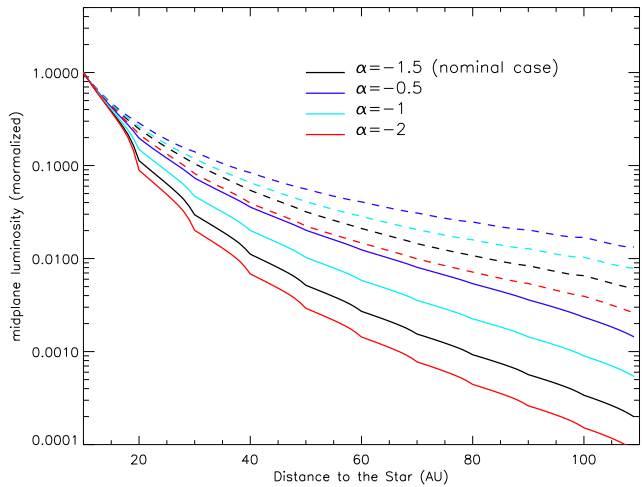}
\caption[]{Normalized mid--plane
surface brightness profiles in scattered light (isotropic scattering), for the 4
different initial density profiles explored in section \ref{sec:param}.
For each case, the full line represents the initial surface brightness profile,
i.e. the profile obtained should the initial $\Sigma \propto a^{\alpha}$
distribution hold for all sizes, and the dashed line shows the final
profile at $10^{7}\,$years.
}
\label{disprof}
\end{figure}
\begin{figure*}
\begin{center}
\includegraphics[angle=0,origin=br,width=0.8\textwidth]{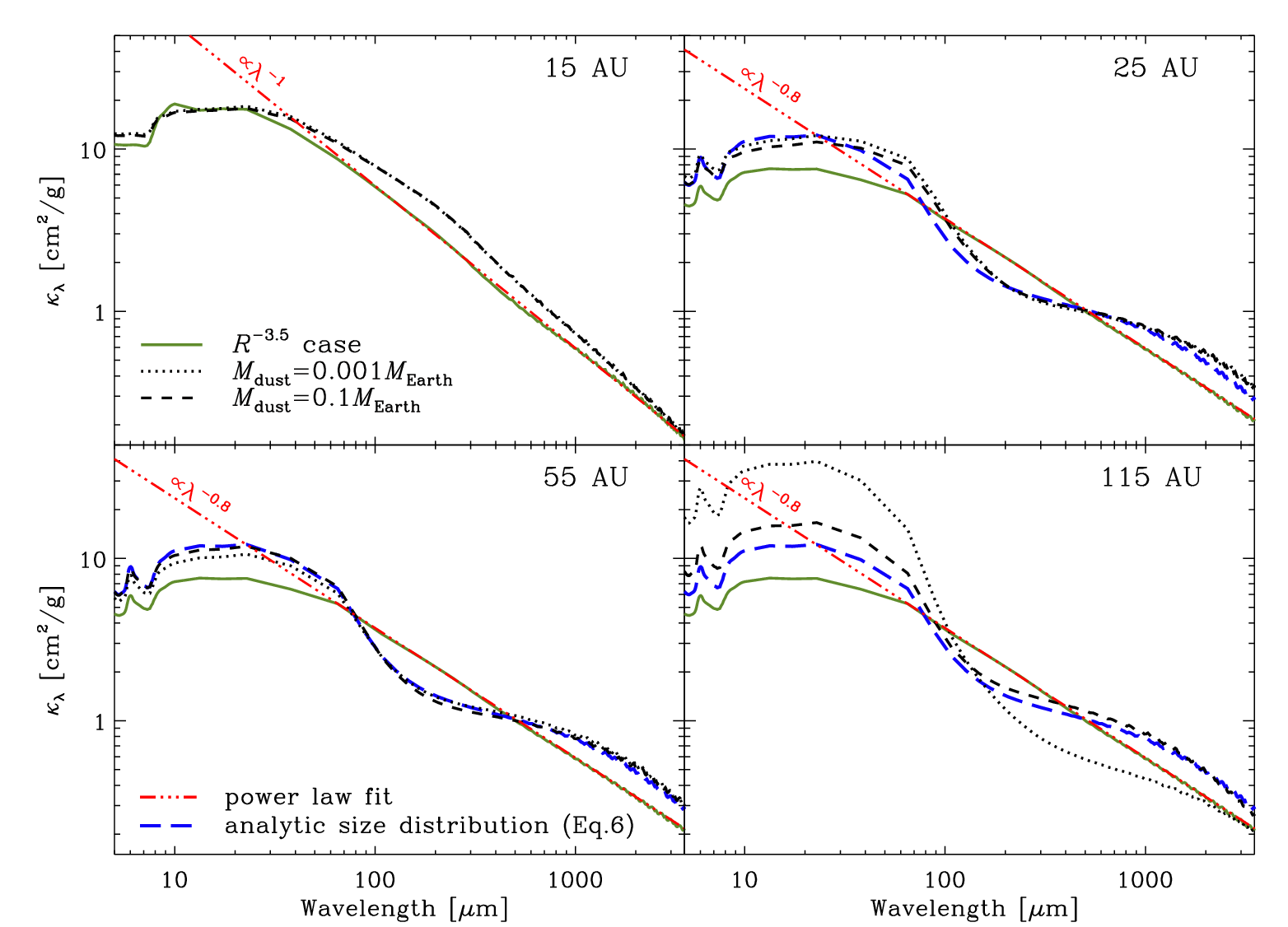}
\caption[]{Mean absorption cross section per unit of mass of material,
$\kappa_{\lambda}$, at $t=10^7$\,yr, for the nominal and low-mass cases
(dashed and dotted line, resp.).
The results are compared to the situation at
$t=0\,$yr ($R^{-3.5}$ size distribution, solid green line), while the red
triple-dot dashed line shows a power law fit to the long-wavelength part
of $\kappa_{\lambda}$.
The mean opacity obtained assuming the fit to the final size distribution
(empirical formula given by Eq.\,\ref{dN}) is overplotted (blue long-dashed line).
 }
\label{kappa}
\end{center}
\end{figure*}

\subsubsection{Comparison to observations}
\label{sec:scatcomp}

Different $S\!B$ profiles are obtained when starting from
different initial radial distributions (cases explored in
Sec.\,\ref{sec:initdens}). However, we see
that while $S\!B\dma{edge}(a)$ profiles do vary with index $\alpha$,
the differences between the initial and final profiles
are remarkably similar, regardless of the initial $\Sigma(a)$ distribution.
Fig.\,\ref{disprof} shows indeed that, for all 4 explored initial $\Sigma(a)$
distributions, the initial $S\!B$ radial profiles always significantly flatten.
The final profiles follow an approximate power law in $S\!B\dma{edge}(a)\propto a^{b}$,
whose index departs from the $t=0$ case by $\Delta b=b-b_0$, with
$\Delta b$ comprised between $-1$ (for the $\Sigma(a) \propto a^{-0.5}$
distribution) and $-1.5$ ($\Sigma(a) \propto a^{-2}$ case)\footnote{Let us
recall that the $S\!B\dma{edge}(a)\propto a^{b_0}$
profile can be interpreted
as the one the system would have in the ``static'' assumption
(as defined in section \ref{sec:profs}), i.e., if an ``equilibrium''
Dohnanyi--like size distribution was to hold and if all particles were to
follow the same spatial distribution as the largest parent--body objects
(whose spatial distribution never significantly departs from the
initial $\Sigma \propto a^{\alpha}$ one).}.
An even more interesting result is that, for a given system,
these final surface brightness profiles $S\!B\dma{edge} \propto a^{b}$
can be directly derived from the
mass surface density distributions $\Sigma(a) \propto a^{\alpha}$
through the following relatively simple approximate
empirical law: 
\begin{eqnarray}
S\!B\dma{edge}(a) \propto a^b \,\,\leftrightarrow\,\, \Sigma(a) \propto a^{\alpha},
{\rm \,\,\,\, with\,\,\,\,} \alpha=2b+3 \,\, .
\label{prof1}
\end{eqnarray}
This relation, valid for isotropic scattering, slightly depends on
the anisotropic scattering parameter $g$. Assuming a \citet{hen41}
phase function, we find:
\begin{eqnarray}
\alpha=2.4b+4.5 & {\rm \,\,\,\, for\,\,\,\,} & \mid g \mid = 0.5 \\
\alpha=3.3b+8.1 & {\rm \,\,\,\, for\,\,\,\,} & \mid g \mid = 0.8 \,\, .
\end{eqnarray}
A useful consequence of these relations is that they 
provide us with a tool to trace back the distribution
of large parent bodies from the observed $S\!B$ profile.
It is important to point out that the distribution
of the \emph{small grains}, those dominating the optical depth, can
still be derived the ``usual'' way from the brightness profiles
(using for example
the $b=\alpha-2$ relation relation for constant opening discs
and grey scattering). The important result is here that recontructing
the optical depth distribution is not equivalent to reconstructing
the mass reservoir distribution.

These results can usefully be compared to the radial luminosity
profiles derived from observations. Although debris discs come in
all sorts and shapes, the general tendency is that most of them
have brightness profiles with a rather steep radial dependence in
$S\!B\propto a^{b}$, with typically $-5<b<-3.5$ for edge-on discs or
$-4<b<-3$ for head-on ones \citep[e.g][]{ard04,gol06,sch06,kal06,kal07}\footnote{we
leave out of this list systems of debris ``rings'' with
razor sharp outer edges, probably sculpted by gravitational
perturbers, like \object{Fomalhaut}, \hr\ or \hdd}. 
These slopes are significantly steeper than the typical $S\!B_{edge}\propto
a^{-2.2}$ obtained for our nominal case with large parent bodies
following the MMSN radial distribution in $a^{-1.5}$. 
From our parameter exploration, only rather extreme cases would lead to
edge-on brightness profiles in $\sim a^{-3.5}$. It would require either
a very steep $\Sigma(a)\propto a^{-4}$ surface density profile
(for the unseen parent bodies) or a very high, and probably unrealistic
anisotropic scattering parameters ($\mid g \mid > 0.9$).
This apparent paradox between our simulation results, which we believe
are rather robust with respect to the flattening of the optical depth
and brightness profiles\footnote{and are moreover confirmed by preliminary
simulations from other teams (Krivov, private communication).}, and observations
might be understood when recalling that our $S\!B_{edge}\propto a^{-2.2}$
profile is obtained \emph{within} the regions where a complete collisional cascade
is assumed to exist, from micron-sized grains all the way up
to big planetesimals. 
There is no obvious reason why the full radial
extents of observed debris discs should correspond to such
collisionnaly active regions.

As a matter of fact, a large fraction of
the luminosity radial profiles of spatially resolved discs could
correspond to regions \emph{outside} the ``parent body'' regions of
collisional activity. For these regions outside the parent body area,
preliminary analytical and numerical results seem indeed to show that
a $a^{-3.5}$ slope could be a typical signature of the presence
of high-$\beta$ grains escaping from their birth region
\citep{stru06,kriv06}\footnote{This outer-edge issue will be addressed
in a forthcoming paper (Th\'ebault \& Wu, in preparation)}.
This possibility is strengthen by the fact that for most debris discs,
the steep slopes are derived in the outer regions located at relatively
large distances from the star: beyond 120\,AU for \bp\,
\citep{gol06}, 130\,AU for \hddd\ \citep{ard04}, $\sim 100\,$AU
for \hd\ \citep{sch06}, $\sim 40$\,AU for \mic\ \citep{kri05},
$55$\,AU for \object{HD\,53143} \citep{kal06}, $\sim 100$\,AU for
\hdddd\ \citep{sch05}. Within the frame of the ``standard''
planet formation scenario, it is likely that these regions are
beyond the limit where accretion of large planetesimals/embryos is
possible \citep[e.g][]{thom03}, so that the presence of
collisional cascades starting from large parent bodies is questionable.
As a consequence, our results imply (within the limitations
to our approach outlined in section \ref{sec:lim}) that an observed
$S\!B_{edge}\propto a^{-3.5}$ luminosity profile is the signature of either:
1) an extended parent body disc with a sharp $\Sigma\propto a^{-4}$ density
decrease (Eq.\,\ref{prof1}) or, more likely, of
2) a region devoid of large particles \emph{beyond} the main disc. 
One robust result is in any case that regions with steady collisional
cascades from large parent bodies, probably cannot result in brightness
profile signatures as steep as $S\!B_{edge}\propto a^{-3.5}$.
Interestingly, for some systems where brightness profiles could be observationally
derived in regions closer to the star, slopes closer to our
nominal $b\sim -2.2$ value have been obtained.
This is in particular true for \bp\, where
in the $\sim$70-100AU region where most of the dust mass is believed to reside,
the brightness profile follows approximately $S\!B_{edge}\propto a^{-2}$
\citep{gol06}.

\subsection{Thermal emission} \label{sec:thermal}

\subsubsection{Dust opacity}
The waviness of the size distribution is well marked for grains smaller
than a few centimetres radius, and should have an observational signature
at far-IR, sub-mm and millimeter wavelengths. The four panels of Figure\,\ref{kappa}
show $\kappa_{\lambda}$, the absorption cross section per unit mass of solid
material, averaged over the size distribution, at four different locations
in the disc. The curves have been obtained assuming spherical grains
made of a silicate core and coated by water ice beyond $20\,$AU
(see Sec.\,\ref{sec:appB1} for more details about
the dust properties).

In the $\d N\propto R^{-3.5}\d R$ case, the mean
opacity $\kappa^0_{\lambda}$ can be approximated by a power law
$\kappa^0_{\lambda}\propto\lambda^{-q}$ beyond $\lambda \sim 70$--$100\,\mu$m,
with $q \simeq 1$ for non-icy grains ($a\la a\dma{sub}=20\,$AU), and
$q\simeq 0.8$ beyond $a\dma{sub}$. These $q$ values compare well with
the theoretical estimates by \cite{dra06}, or the best fit  values
obtained for debris discs \citep[e.g.][]{den00,gre04}.
Nevertheless, Fig.\,\ref{kappa} shows that realistic collisional systems
do have mean opacities that strongly depart from a simple power law profile
at long wavelengths. At representative distances from the star ($25$\,AU and
$55\,$AU), the mean opacity $\kappa_{\lambda}$ shows a characteristic dip at
$\lambda \sim 150$--$200\,\mu$m, and a bump at millimetre wavelengths for both
the nominal and low-mass cases. At $55\,$AU for example,
the mean opacity ratio in the {\it Spitzer}/MIPS2 and MIPS3 bands, $\kappa_{70\,\mu\mathrm{m}}
/ \kappa_{160\,\mu\mathrm{m}}$, amounts to $1.7$--$2$
times the mean opacity ratio should a $R^{-3.5}\d R$ size distribution
hold. Similarly, $\kappa_{520\,\mu\mathrm{m}} / \kappa_{160\,\mu\mathrm{m}}$,
$\kappa_{850\,\mu\mathrm{m}} / \kappa_{160\,\mu\mathrm{m}}$, $\kappa_{1300\,\mu\mathrm{m}}
/ \kappa_{160\,\mu\mathrm{m}}$, are $1.7$, $2.1$ and $2.4$, respectively, larger than those
found for a Dohnanyi size distribution.

In Sec.\,\ref{sec:fitsize}, we provide an anlytical fit to the final
size distribution responsible for the waviness of the mean opacity. The mean
opacity obtained assuming the empirical differential size distribution
given by Eq.\,\ref{dN}, compares well to that calculated at our representative
distance of $55$\,AU.

\subsubsection{Disc SED and images}
\begin{figure}
\includegraphics[angle=0,origin=br,width=\columnwidth]{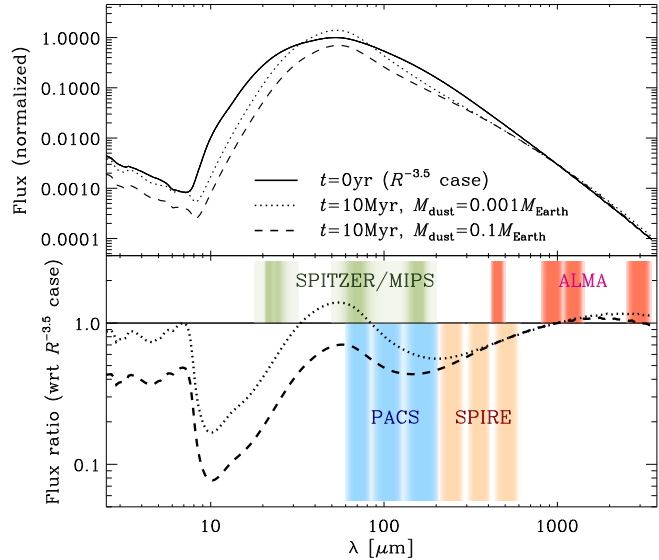}
\caption[]{Disc SEDs, including both thermal emission and scattered light (which
dominated over thermal emission at $\lambda \la 10\,\mu$m).
 }
\label{SEDs}
\end{figure}

The actual impact on the disc spectral energy distribution (SED) is displayed
in the top panel of Fig.\,\ref{SEDs}, where the synthetic SEDs have been calculated
using the model of \citet{aug99}. The solid line on the figure represents
the disc SED, normalized to $1$ at its maximum, at $t=0$ ($\d N \propto R^{-3.5} \d R$
size distribution), and the bottom panel shows the flux ratio
after $10\,$Myr of evolution of the system.
As anticipated, the wavy structure of the size distribution has an observational
counterpart at far-IR to millimetre wavelengths, and in particular a lack of emission
in the $150$--$200\,\mu$m spectral range compared to the $R^{-3.5}$
size distribution. The predicted disc colors depart from the
Dohanyi case by factors that compare to the mean opacity ratios calculated above.
More precisely, the $70\,\mu$m to $160\,\mu$m, $520\,\mu$m to $160\,\mu$m,
$850\,\mu$m to $160\,\mu$m, and $1300\,\mu$m to $160\,\mu$m flux ratios,
are $1.5$--$2.1$, $1.3$--$1.8$, $1.6$--$2.2$ and $1.9$--$2.4$,
respectively, larger than those found for a Dohnanyi size distribution.

The $150$--$200\,\mu$m spectral range clearly appears as a critical spectral range
to test the model developped in this paper. It requires a good sampling of the
SED at long wavelengths, and a sufficiently precise relative photometric
calibration. Several observational facilities working at far-IR to millimeter
wavelengths, will start operation in a very near future. Some of them
are indicated in the bottom panel of Fig.\,\ref{SEDs}, to which
should be added the SCUBA-2 camera at JCMT \citep{hol06}, and the SOFIA observatory
\citep{bec06,cas06}. The PACS and SPIRE instruments onboard the Herschel space
observatory are particularly well suited to identify the dip at around $150$--$200\,\mu$m
by measuring the exact shape of debris discs SEDs beyond $\lambda \sim 70\,\mu$m
\citep{pil05,pog06}. This would allow to find a direct observational signature of
an ongoing collisional cascade in a debris disc. 

\begin{figure*}
\includegraphics[angle=0,origin=br,width=\textwidth]{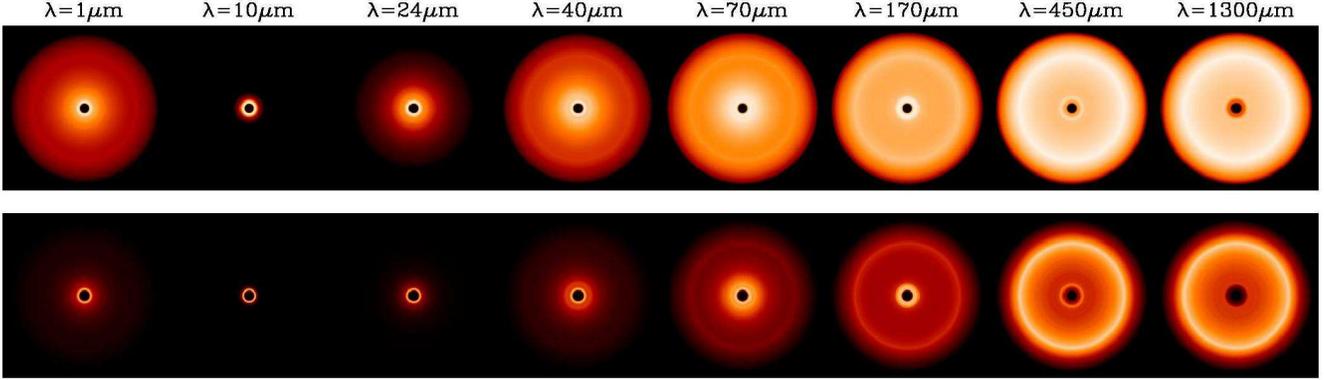}
\caption[]{Appearance of the high--mass disc, assumed face-on, at $t=10\,$Myr as
a function of the observing wavelength. 
{\it Top panel}: logarithmic scale. {\it Bottom panel}: linear scale.}
\label{images}
\end{figure*}

The dependence of the size distribution on the distance to the star,
evidenced in Figs.\,\ref{surfcomp} and \ref{opt}, has direct consequences
on the appearance of the disc, as illustrated in Fig.\,\ref{images}.
In the near-infrared, and at shorter wavelengths,
light scattering by small (high--$\beta$) particles dominates the disc image.
The disc therefore shows a decreasing brightness profile with increasing $a$
as discussed in Sec.\,\ref{sec:profs}.
In the thermal emission-dominated regim (mid-infrared and beyond),
the disc morphology totally depends on the observing wavelength.
At $\lambda = 24\,\mu$m for instance, the disc surface brightness smoothly decreases
with the distance from the star, while at (sub-)mm wavelengths, the disc shapes
a ring peaked close to the outer edge of the parent-body disc ($\sim 100\,$AU),
a situation that interestingly recalls the case of the Vega disc \citep{su05}.

\section{Empirical formulae for debris disc modeling}
\label{sec:discu}

The purpose of the present work is to numerically explore
the collisional evolution of an extended debris disc, when taking into
account the crucial effects of impacts induced by the radiation--pressure
affected small grains. The different results displayed in Secs.\,\ref{sec:param} 
and \ref{sec:obs} show that, although noticeable differences
might be observed for different setups, important generic trends
can be derived. We propose, in the following, empirical laws for the
size distribution and collision timescales, that can be used for debris
disc modeling as alternatives to the classical $R^{-3.5}$ size distribution
and to the $t^0\dma{coll}=\left(\tau \Omega\right)^{-1}$ law.

\subsection{Fit to the size distribution}
\label{sec:fitsize}

\begin{figure}
\includegraphics[angle=0,origin=br,width=\columnwidth]{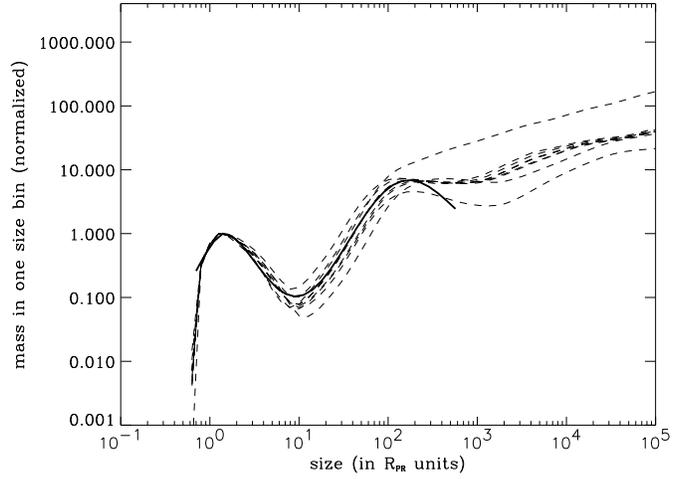}
\caption[]{{\it Dashed lines}: Final size distribution profiles,
averaged over the whole system,
for all numerically tested cases, except the 2 ``weak'' and ``hard''
material cases with different collisional prescriptions. 
All profiles' $x$--axis have been renormalized to units of $R\dma{PR}$, and
all $y$--axis to the value of the first wavy ``peak'' at $R\simeq 1.5R\dma{PR}$.
{\it Solid line}: reference profile derived from our empirical fit
given by Equ.\,\ref{dM} and integrated over one logarithmic size interval:
$\Delta M(R) = G(R) \left(R/1.5R\dma{PR}\right)^{0.41} \Delta M_{(1.5R\dma{PR})}$.
}
\label{profnorm}
\end{figure}
One crucial result concerns the final size distributions.
For almost all runs, the system always quickly reaches a
quasi steady--state, with a pronounced wavy distribution 
which strongly departs from a standard
``equilibrium'' distribution in $\d N\propto R^{-3.5}\d R$, or any
simple power law in $R^{q}\d R$ for that matter.
As clearly appears in Figs.\,\ref{snapsh} \& \ref{snapslmc}, the waviness
varies with location in the system, it is less pronounced close
to the inner edge, since it is mostly due to collisions due 
to high velocity outward moving small grains.
However, if ones considers the average distribution
integrated over the whole disc, we have seen that its profile
only weakly depends on parameters such as
the system's total mass, it's dynamical excitation
or the value of the radiation pressure cut--off size $R\dma{PR}$. 
For the latter case,
what is observed is mostly an offset of the wavy--distribution,
which retains its global shape and main characteristics. As for
the total initial mass, it does not crucially affects
the final $shape$ of the size distribution as long as collision lifetimes
of dust grains are shorter than the system's age (see Sec.\,\ref{sec:lowm}).
The final size distribution is even relatively unaffected by the
profile of the initial mass distribution (exponent of the $\Sigma(a)$
profile). 
The only cases for which a major modification of the size
distribution  is observed are the ``very weak'' and the
``very hard'' material cases.
Apart from these 2 exceptions, for all other 10 tested setups
we obtain very similar features:
a strong depletion of $R\la R\dma{PR}$ grains, a peak
for $R\simeq 1.5R\dma{PR}$ followed by a deep depletion
of objects in the $10R\dma{PR}<R\la 50R\dma{PR}$ range.
The similarities between all profiles are even more striking when
they are renormalized by their value at $R=R\dma{PR}$
(Fig.\,\ref{profnorm}). As can be clearly seen, variations are
very limited for $R\la 100R\dma{PR}$. For this size range it seems
thus reasonable to consider that, as a first approximation, 
the size distribution obtained in our nominal case is
a relatively good standard for spatially extended systems.
We were able to derive an empirical fit for this revised
size distribution, valid in the $R\la 100R\dma{PR}$ range. When written
in terms of the differential mass distribution, it reads:

\begin{equation}
\d M \propto\,G(R) R^{-0.59} \d R
\label{dM} 
\end{equation}
with
\begin{equation}
\log_{10}{\left(G(R)\right)}=\frac{2}{3} 
\left[ \cos \left( 2\pi \left[ \left| \frac{1}{2}\log_{10}\left(\frac{R}{1.5 R\dma{PR}}\right) \right| \right]^{0.85}\right) - 1  \right] \, .
\end{equation}
This new relation proves
to be a reasonably good fit to almost all profiles in the $R\la 100R\dma{PR}$ range
(Fig.\,\ref{profnorm}). In terms of the differential size distribution
$\d N(R)$, this translates into
\begin{equation}
\d N \propto\,G(R) R^{-3.59} \d R \, , \,\, \mathrm{for }\,\, \frac{2}{3}R\dma{PR} < R \la 100\,R\dma{PR} \, .
\label{dN} 
\end{equation}
Beyond $100\,R\dma{PR}$, stronger divergences between different runs are observed.
However, as a rough first order approximation, 
the differential size distribution can approximately be extrapolated by
a $R^{-3.7}$ power law. The $R^{-3.7}$
extrapolation has been used to calculate the mean opacity represented by a
blue long-dashed line in Fig.\,\ref{kappa}.

\subsection{Fit to the collisional particle lifetime}

\begin{figure}
\includegraphics[angle=0,origin=br,width=\columnwidth]{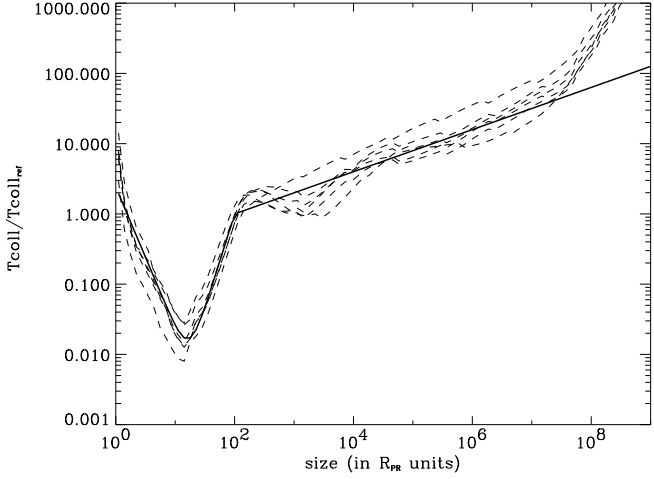}
\caption[]{{\it Dashed lines}: Collisional lifetimes at $a_0=55$AU, for all
tested cases except the weak and hard material runs, when renormalized
by the reference timescale $t\dma{coll-ref}= t^0\dma{coll}(a_0)
\left(\Omega(a_0)\tau(a_0)\right)^{-1}$.
{\it Solid line}: profile derived from our empirical fit
(Equs.\,\ref{tcfit1} and \ref{tcfit2})
}
\label{Tcnorm}
\end{figure}

As shown in Sec.\,\ref{sec:colllife}, collisional lifetimes strongly
vary with particle sizes: they increase very rapidly when $R$ gets
close to the blow--out limit $R\dma{PR}$, reach a sharp minimum
around $R\simeq 10R\dma{PR}$, increase sharply again between
$10R\dma{PR}$ and about $100R\dma{PR}$ and then continue to increase
much more slowly with increasing sizes (see Fig.\ref{tcoll}).
We have also shown that a direct consequence of this result is that
collisional lifetimes cannot be directly derived from the
optical depth through the simplified formula $t^0\dma{coll}(a)=(\tau\Omega)^{-1}$.
There are several reasons why this formula cannot hold here:
\begin{itemize}
\item
the $t^0\dma{coll}(a)=(\tau\Omega)^{-1}$ formula implicitly considers impacts
between objects of {\it equal} sizes, thus neglecting the broad size spectra of
all possible impactors on a given target,
\item
it also implicitly assumes that all impacts are
fully destructive, i.e., that the collision timescale is equal
to the collisional {\it lifetime}. This neglects all cratering
impacts, whose role is crucial for the considered problem
(see Fig.\ref{cutcra}),
\item
even more important: this formula neglects all effects due
to the specific dynamics of the smallest grains affected by
radiation pressure,
\item
last but not least: at any given distance $a_0$ from the star,
it neglects all collisions due to objects coming the inner
$a<a_0$ regions, and it has been shown (Fig.\ref{cutcra}) that
these collisions are crucial for the evolution of dust grains.
\end{itemize}

As can be seen for example in Figs.\,\ref{tcoll}a,b and \ref{tcolle}a,b, 
collisional timescales significantly vary for different initial
conditions, in particular total initial mass and dynamical excitation
of the system. However, the {\it profiles} of the $t\dma{coll}(a,R)$
curves are relatively similar. In order to visualize
these similarities more clearly, all $t\dma{coll}(a,R)$ curves
have been renormalized by
the reference timescale $t^0\dma{coll}(a)=(\tau\Omega)^{-1}$ (Fig.\ref{Tcnorm}).
In a similar fashion as for the size distributions, we
see that all normalized $t\dma{coll}(a,R)$ profiles remain relatively close to
the nominal case\footnote{The significant differences observed between the dynamically
excited and dynamically cold cases (see Fig.\ref{tcolle}) are
partially erased after renormalization by $t^0\dma{coll}(a)$. Indeed, as
seen in Fig.\ref{excomp}, systems with low $\langle e\rangle$ are globally depleted
in $R\la0.1$mm grains and have thus lower optical depth (since $\tau$
is mostly contained in the smallest particles)}.
This opens the possibility for deriving an empirical fit to
$t\dma{coll}(a,R)$ as a function of $a$ and $\tau$:
\begin{equation}
t\dma{coll}(a,R)= t^0\dma{coll}(a)\,\left[\left(\frac{R}{R_1}\right)^{-2}
+\left(\frac{R}{R_2}\right)^{2.7}\right] \,\,\, {\rm for}\,\,\,  R<R_2
\label{tcfit1} 
\end{equation}
with $R_1=1.2R\dma{PR}$ and $R_2=100R\dma{PR}$, 
and
\begin{equation}
t\dma{coll}(a,R)= t^0\dma{coll}(a)\,\left(\frac{R}{R_2}\right)^{0.3}
\,\,\,\,\,\,\,\,\, {\rm for}\,\,\,  R>R_2 \,\, .
\label{tcfit2} 
\end{equation}

\subsection{Approximations and limitations}
\label{sec:lim}

Let us state again that these relations should be taken with care.
An important general remark is to again stress
that they have been derived for extended collisionally active regions, i.e.,
regions with steady collisional cascades starting from large
reservoirs of big unseen parent bodies. These regions might not
account for all the observed radial extent of debris discs: some
observed regions are probably collisionally inactive areas where only
small high-$\beta$ grains, produced in parent body regions further
inside, are present (see discussion in Sec.\,\ref{sec:scatcomp}).

Moreover, within the frame of our numerical approach it is important to point
that these fits are valid for our nominal collision outcome prescription, and that
significant variations should be expected for harder or weaker
material prescriptions (Fig.\,\ref{compcoll}). 
It should also be noted that in a ``real'' disc,
all individual particles are not completely identical: they would
have slightly different material compositions, porosities, differ in presence or
absence of microcracks, etc... This might alter the size distribution profile,
probably damping the waviness described in Eq.\,\ref{dN} to some extent,
but such sophisticated effects are difficult to take into account
with a particle-in-a-box code.
Another important point is the fact that the smallest particles considered
here are just below $R\dma{PR}$, so that only 2 size ``bins'' correspond
to unbound so-called ``$\beta$-meteoroids''. We
nevertheless performed a few test runs with additional
small-size bins, and observed no drastic change in the final profiles.
However, for more massive discs, taking into
account the role of $\beta$-meteoroids, as was done
in the pioneering work of \citet{kriv00}, might be crucial. For such high-mass
systems, extremely efficient collisional ``avalanches'' chain reactions
triggered by $\beta$-meteoroids could possibly play a significant role
\citep{grig07}. The contribution of unbound grains could
also be important for interpretation of observations
particularly sensitive to smaller particles, e.g. polarimetry \citep{kriva00}.

We do however believe that, regardless of their exact level of accuracy,
the present empirical fits are in any
case a more reliable fit to ``real'' size distributions than any
simple $\d N\propto R^{q}\d R$ power law (be it $q=-3.5$ or not) extrapolation.

\section{Summary and conclusions}
\label{sec:conclu}

We elaborate in this paper a model able to follow the collisional
evolution of extended debris discs over a $10$\,Myr span. We confirm
the previous results obtained by \citet{theb03} for a narrow, isolated
annulus, that the classical $\d N \propto R^{-3.5}\d R$ Dohnanyi size
distribution cannot hold in realistic collisional discs. Rather, a
wavy size distribution develops in the whole system, amplified
by the particular dynamics of the radiation pressure affected grains
(high-$\beta$ particles).

The model builds on the classical particle-in-a-box technique, and allows
a detailed exploration of the various parameters that impact the disc
evolution. Such a quantitative numerical exploration had not been undertaken so far,
at least not when following the size distribution evolution
over a range encompassing all objects from
the $\mu$m to the biggest
parent bodies in the 50\,km range\footnote{with the notable exception
of the very innovative and promising kinetic approach of \citet{kriv06},
but so far considering a very simplified model of collision outcomes}.
We chose therefore not to focus on 
one given observed debris disc but to consider
a fiducial nominal system, making the most reasonable
(or maybe least unreasonable) assumptions, in order to
clearly identify and quantify the complex mechanisms at play,
and derive general behaviours without biases by non--generic artifacts. 
However, in order to check the robustness of our results,
several key free parameters have been explored.
Our main results can be summarized as follows:
%
\begin{enumerate}

\item A wavy size distribution, strongly departing
from a $\d N \propto R^{-3.5}\d R$ power law,
is a common feature of collisional debris discs. 

\item The wavy pattern includes an overdensity of grains with
radius about twice the blow-out grain size $R\dma{PR}$, and a strong
depletion of the $10-50\,R\dma{PR}$ particles.

\item The waviness weakly depends on the disc mass, initial surface
density profile, mean disc dynamical excitation, stellar properties, but is affected
by the collision outcome prescription, especially the resistance of
objects to collisions.

\item In extended discs the evolutions of different regions of the systems are
strongly interconnected: the waviness is amplified by high-$\beta$ bound
particles (grains strongly affected by pressure forces), which have large radial
excursions within the system and can impact, at very high velocities,
larger objects far outside
the region where they were initially produced. 

\item Surprisingly, the global dust to planetesimal mass ratio is,
to a first order, not strongly affected by the size distribution waviness.

\item Collisional lifetimes strongly differ
from the usual $(\tau \Omega)^{-1}$ approximation in realistic collisional
systems.

\item The optical depth and the scattered light flux are dominated by a very narrow
range of so-called $\alpha$-meteoroids, i.e., bound objects just
above the blow--out cutoff size.

\item {\it Spatial} distributions are also affected.
The radial distributions of grains of different sizes might significantly
diverge from one another. More generally, there is a major
discrepancy between the radial distribution of particles in
the dust--size range, i.e. those accessible to observations,
and the largest undetectable objects that make up most of the system's mass.
The distribution of small grains, and thus of the disc's optical
depth, is significantly flatter than that of the big parent bodies.

\item This flattening of the small grains radial
distribution translates into a flattening of
surface brightness profiles in scattered light in the regions
where the big parent bodies reside.
For a disc having an initial MMSN surface density profile
the equilibrium scattered light surface brightness profile is roughly
in $S\!B_{edge}\propto a^{b}$, with $-2.3<b<-2$ instead of the standard
$b\simeq -3.5$ value.

\item These radial slopes are less steep than those observed for 
the vast majority of debris discs. This apparent
paradox could be explained by the fact
that for most systems, radial brightness profiles are
observed in regions \emph{beyond} the outer edge of the main
``parent body'' disc. In these regions, no collisional cascades take place
and only small high-$\beta$ grains, produced further inside and pushed
on eccentric orbits by pressure forces, are observed.

\item The waviness of the size distribution translates into wavy dust opacities
and SEDs at far-IR and (sub-)millimeter wavelengths, which could be
observable signatures of the collisional activity in debris discs.

\item We derive an empirical formula for the differential size
distribution (Eq.\,\ref{dN}) which fits reasonably well the numerically
obtained results. Although this approximate fit should be taken
with care because of the unavoidable limitations of our numerical code,
future models aiming at reproducing multi-wavelength 
observations might use this formula as an alternative
to simplified $dN\propto R^{q}dR$ power laws.

\item Similarly, we propose an empirical formula for the collisional
lifetime of the particles (Eq.\,\ref{tcfit1} \& \ref{tcfit2}) that
might be used to interpret data.

\end{enumerate}

This paper provides the basis for future debris discs modeling of
individual cases such as \object{Vega}, for which both resolved data and numerous
photometric measurements are available. But overall, the waviness of the size
distribution is becoming a well established
feature that cannot be ignored in future SED analysis, and the empirical
size distribution given by Eq.\,\ref{dN} is provided for this purpose.
We in addition stress that a wealth of future facilities working at far-IR
and (sub-)millimeter wavelengths (Herschel, SOFIA, SCUBA-2, ALMA) will soon
offer the opportunity to test the model developed in this paper, providing
a direct observational hint for an ongoing collisional cascade in a debris
disc.

\begin{acknowledgements}
The authors thank the reviewer Alexander Krivov for very useful comments
that helped significantly improve the paper.
We also thank Patrick Michel for fruitful discussions
on collision outcome prescriptions.
This work was partly supported by the European Community's
Human Potential Program under contract HPRN-CT-2002-00308,
PLANETS.
\end{acknowledgements}

{}
\Online
\begin{appendix}


\section{Evolution equation}
\label{sec:appA}

The present multi--annulus code is based on the single--annulus algorithm
developed in TAB03 and described at length in this paper.
We shall thus only recall here its main characteristics before describing in
more details the enhancement performed for the present version.

We first spatially divide the system into $N\dma{a}$ concentric annuli.
Within each annulus, we follow the
classical particle--in--a--box approach in which the particle
population is divided into $n$ boxes each standing for a given particle size
$R\dma{i}$.
In each annulus $ia\geq2$ (all except the innermost one), additional bins
are included which account for the small grains
originating from $ia\arcmin<ia$ annuli and placed by radiation pressure
on highly eccentric or unbound orbits crossing the $ia$ annulus.
We arbitrarily set the limit for grains sizes for which additional bins
are considered by the criteria $\beta_i>\beta\dma{lim}=0.05$.
For one given particle size $R\dma{i}$ in the $ia$ annulus, there are thus
$1+n_{b(i)}$ corresponding bins, where $0\leq n_{b(i)}\leq ia$ is the
number of possible source annuli $ia\arcmin<ia$ for all ``foreign born''
$R\dma{i}$ grains. To describe the number of particles of one given grain population
within one given annulus $ia$, we use the terminology
$N_{ia,i,ia'}$, where $1\leq i\leq n$ is the size bin index and $ia'\leq ia$ the
source annulus where the grain population has been produced ($ia'=ia$
for all particles with $\beta_i<\beta\dma{lim}$).

At each time step, the change in the number $N_{ia,i,ia'}$
is given by the collisional
evolution equation displayed in Equ.1 of TAB03. 
For the $locally$ produced small grains affected by radiation pressure
($\beta_{i}>\beta\dma{lim}$), an additional term is introduced which reads

\begin{equation}
dN_{ia,i,ia} = - f_{out(ia,i,ia)}\,N_{ia,i,ia} +
f_{in(ia,i,ia)}\,N_{ia+1,i,ia}
\label{dn1}
\end{equation}
where $f_{out(ia,i,ia)}$ is the fraction of $N_{ia,i,ia}$ particles leaving
the $ia$ for the $ia+1$ annulus during $dt$ and $f_{in(ia,i,ia)}$ the fraction
of particles re--entering the $ia$ annulus after completing
one full orbit ($f_{in(ia,k,dt)}=0$ for grains on unbound
orbits). Note that these re--entering particles necessarily come from the 
neighbouring $ia+1$ annulus, where they were members of the
$N_{ia+1,i,ia}$ bin. All
$f_{out(ia,i,ia)}$ and $f_{in(ia,i,ia)}$ rates are derived from
separate deterministic numerical simulations following the dynamical
behaviour of 10000 test particles with $\beta=\beta_{i}$ released
on randomly distributed orbits with $e=2i=\langle e\rangle_{0}$ and
$a\dma{min}=10AU<a<a\dma{max}=120AU$.

For the $\beta_{i}>0.05$ grains which have been originally produced
in an inner $ia'<ia$ annuli, the additional
evolution term due to inter-annuli exchanges reads

\begin{eqnarray}
dN_{ia,i,ia'} \nonumber \\
- g_{out+(ia,i,ia')}\,N_{ia,i,ia'} - g_{out-(ia,i,ia')}\,N_{ia,i,ia'}
\nonumber \\
+ g_{in+(ia,i,ia')}\,N_{ia+1,i,ia'} + g_{in-(ia,i,ia')}\,N_{ia-1,i,ia'}
\label{dn2}
\end{eqnarray}
where $g_{out+(ia,i,ia')}$ and $g_{out-(ia,i,ia')}$ are the fraction of
outgoing (to the $ia+1$ and $ia-1$ annuli respectively) particles
and $g_{in+(ia,i,ia')}$ and $g_{in-(ia,i,ia')}$ the
fraction of incoming (from the $ia+1$ and $ia-1$ annuli) particles.
The terms $g_{in-(ia,i,ia')}$ and $g_{out+(ia,i,ia')}$ correspond
to particles produced in the $ia'$ annulus on their way out towards
their apoastron (or infinity for unbound orbits) and
the terms $g_{in+(ia,i,ia')}$ and $g_{out-(ia,i,ia')}$ correspond
to particles having already reached their apoastron and on their
way back to the $ia'$ annulus (these terms are equal to zero for
unbound orbits). All 4 parameters are estimated with the same type of
numerical simulations as those used for deriving 
$f_{out(ia,i,ia)}$ and $f_{in(ia,i,ia)}$.

As already mentioned, the dynamical state of the system is fixed and does not
evolve with time. To estimate the average encounter velocities,
we divide all possible target-impactor encounters into two types: 1) those
involving two $\beta_{i}<0.05$ (i.e. not significantly affected by radiation
pressure) particles, and 2) those where at least one of the involved bodies
is on a radiation pressure modified orbit ($\beta_{i}>0.05$)
For type 1) impacts within one given annulus $ia$ at a distance $r_{ia}$
from the star, the encounter velocity is simply given by the classical
expression \citep[e.g.][]{lisste93,theb03}:

\begin{equation}
\langle dv\rangle_{i,ia;j,ia} = \left( \frac{5}{4} \langle e^{2}\rangle + \langle
i^{2}\rangle \right)^{1/2}\,\langle v_{kep(ia)}\rangle
\label{dv}
\end{equation}
where $\langle v_{kep(ia)}\rangle $ is the average Keplerian velocity at
distance $r_{ia}$, and $\langle e\rangle $
and $\langle i\rangle $ are the average orbital parameters imposed as initial
conditions. As described in section 2, we take here
$\langle e\rangle_{0} =0.1=2\,\langle i\rangle_{0}$.
For type 2) impacts the average impacting speed are numerically estimated
in specific determinisitic numerical runs.

\section{Collision outcomes}
\label{sec:appB}

We follow here the classical approach where collision outcomes are divided
into 2 regimes, depending on the ratio between the specific impact
energy per target mass unit $Q\dma{imp}=E\dma{col}/M\dma{t}$, where
$E\dma{col}=M\dma{p}M\dma{t}\Delta v^{2}/2(M\dma{p}+M\dma{t})$,
and the critical specific energy $Q^{*}$:
{\it catastrophic fragmentation} if $Q\dma{imp}>Q^{*}$ and
{\it cratering} if $Q\dma{imp}<Q^{*}$.

\subsection{Critical specific energy}
\label{sec:appB1}

$Q^{*}$ is a function of the target's radius $R_t$, this dependence being
usually expressed as the combination of two power laws \citep[e.g.][]{benz99}:

\begin{equation}
Q^{*}\dma{(R_t)} = Q_{0s}\left(\frac{R_t}{1cm}\right)^{b_s} +
B\rho\left(\frac{R_t}{1cm}\right)^{b_g}
\label{Q}
\end{equation}
The first term on the right hand side corresponds to the $strength$ regime,
valid for small sizes, where $Q^{*}$ slowly decreases with size, while the second term
has a positive index corresponding to the $gravitational$ binding regime.
Values for $Q_{0s}$, $B$, $b_s$ and $b_g$ depend on the physical composition
of the objects and are derived from laboratory experiments or
numerical models \citep[e.g.][]{hh90,dav90,hol94,paol96,benz99,ara99}.

\begin{figure}
\includegraphics[angle=0,origin=br,width=\columnwidth]{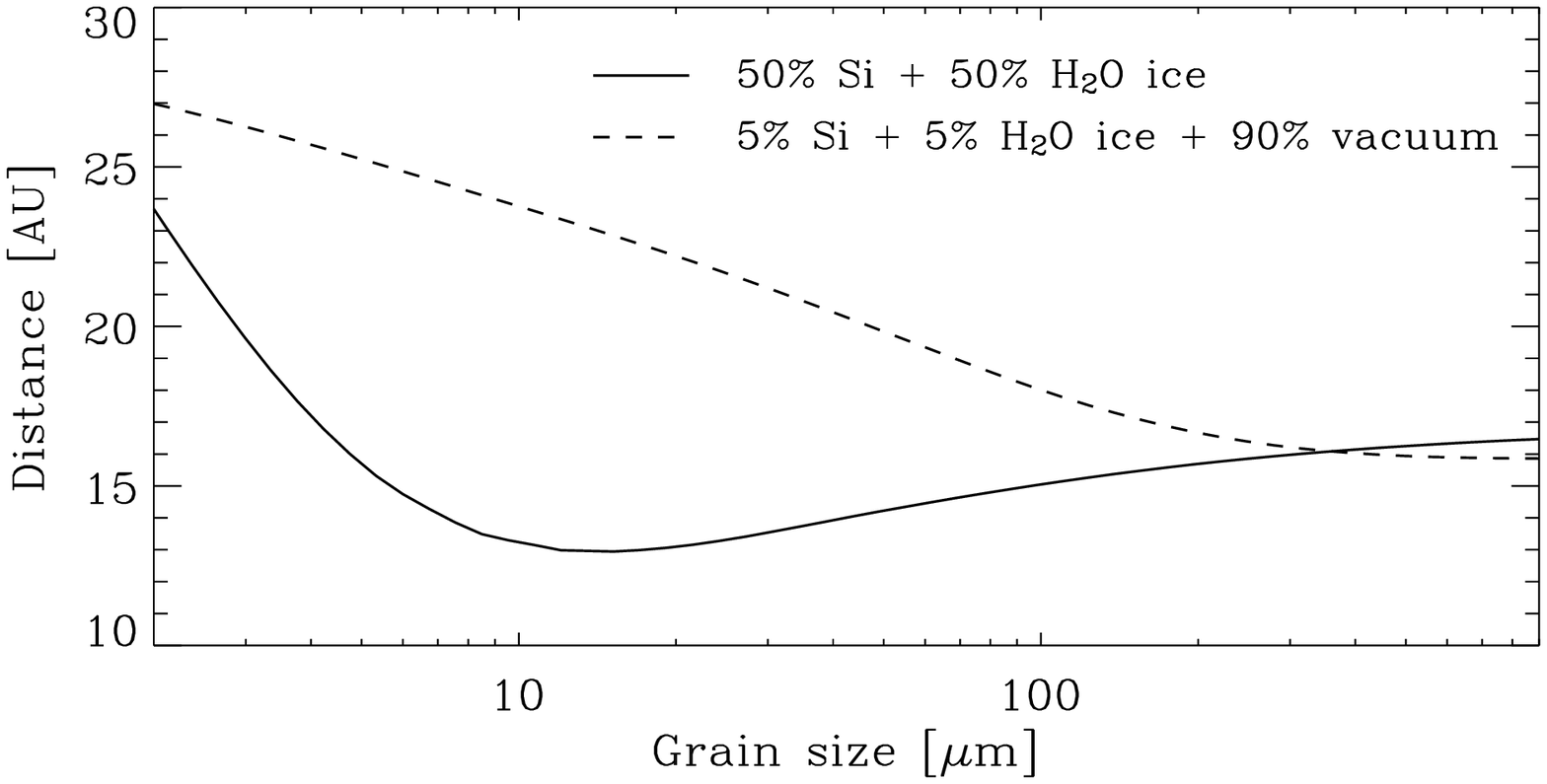}
\caption[]{Position of the snow line ($T\simeq 120\,$K) for grains
with silicate-rich cores around a \bp-like star. A NextGen synthetic
stellar atmosphere spectrum for a A5V star \citep{hau99} has been
used to compute the equilibrium temperature of the grains. The silicates
and water ice optical constants are from \citet{dra03} and \citet{ligre98},
respectively. The mean grain optical index was obtained using
the Maxwell-Garnett mixing rule, and the Mie theory used to compute
the grains absorption/emission efficiencies.}
\label{dTsub}
\end{figure}
A first important issue is then which chemical composition
to assume for the objects: proportion of ices and silicates,
porosity, etc... For our nominal case of a A5V star, we have
assumed simple mixtures of silicates, water ice, and vacuum
(to mimic porosity). The ice sublimation distance is a function
of the grain size, as shown in Fig.\,\ref{dTsub} for
two different vacuum volume fractions ($0$\% to simulate
non-porous grains, and $90$\% for highly porous grains).
For grains larger than a few $\mu$m, the sublimation distance oscillates
about $20\pm5$\,AU, with the largest distances
reached for the smallest grain sizes considered in this paper.
Given the spatial resolution of our simulations, we adopt a single,
average sublimation distance for water ice of $\sim 20$\,AU.

But the problems do not stop here since, even for similar materials,
the different $Q^{*}$ prescriptions available in the literature
do often significantly diverge from one another
\citep[see for example Fig.8 in][]{benz99}.
For icy bodies in particular, critical energy estimates might differ
by up to two orders of magnitude depending on the studies
\citep[see for example the discussion in section 4 of][]{bur05}. 
It is not the purpose of the present work to address this very difficult
issue.

We shall here consider a ``nominal'' case, assuming that for silicates
the critical energy is the one derived by \citet{benz99} for impacts at
3\,km.s$^{-1}$, i.e., $Q_{0s}=3.5\times 10^{7}$erg.g$^{-1}$, 
$B=0.3\,$erg.cm$^{3}$.g$^{-2}$, $b_s=-0.38$ and $b_g=1.36$. For ices,
we follow \citet{kriv06} and take
$Q^{*}\dma{ice(R_t)}=1/5\,Q^{*}\dma{sil(R_t)}$, which
is an approximate intermediate position between the hydro--code
results of \citet{benz99}, who found that ice could be almost as resistant
as silicates, and most experiment results in which ices proved
to be more than one order of magnitude weaker \citep[see][and
reference therein]{bur05}.
Because the smaller radiation--pressure affected grains might impact
objects in regions different from the ones where they have been
produced, we have also to take into account impacts where one colliding body
is icy and the other one rocky. For these heterogeneous collisions,
we assume a simple prescription with $Q^{*}\dma{ice-sil}=1/2Q^{*}\dma{ice}$
and $Q^{*}\dma{sil-ice}=2Q^{*}\dma{sil}$, roughly taking into account the fact that
impacts by hard (resp. weak) projectiles on weak (resp. hard) targets
are more (resp. less) destructive than impacts between bodies of same material.
Furthermore, since impacting velocities do significantly vary within the disc, 
we take into account the weak $Q^{*}$ dependence on $\Delta v$ found
by \citet{hh90} and assume

\begin{equation}
Q^{*}_{(\Delta v)} = Q^{*}_{(3.km.s^{-1})}\,\left(\frac{\Delta v}
{3{\rm km.s}^{-1}}\right)^{0.35}
\label{Qv}
\end{equation}

Note that our critical energy prescription gives $Q^{*}$ values
significantly higher than those of \citet{kriv06}. These authors assumed
$Q_{0s}=3\times10^{6}$erg.g$^{-1}$ (at R=1cm) for silicates, which is 
significantly below most $Q_{0s}$ estimates available in the literature
\citep[see Fig.8 of][]{benz99},
with the exception of that of \citet{dur98} obtained from
fitting the observed size--distribution of asteroids.
The prescription of \citet{kriv06} will however be tested as
a ``weak material'' run (see section \ref{sec:collpres}). 

\subsection{catastrophic fragmentation}
\label{sec:appB2}

If $Q\dma{imp}>Q^{*}$ {\it catastrophic fragmentation} occurs:
the target is shattered and
produces a population of fragments where the biggest one has a mass
$M\dma{lf}<0.5\,M_t$. The value of $M\dma{lf}$ as well as the size distribution of
the produced fragments is computed following the procedure described at length in
section 2.4 of TAB03. 

We would like to point out that our model departs from
the often assumed simplifying assumption that fragment size distribution follows
a power law in  $dN \propto R^{-3.5}dR$ \citep[e.g.][]{aug01,kriv06}.
Such a power law is indeed in principle the $equilibrium$ value reached 
after sufficient mutual collisions, but not the one for fragments
produced after $one$ given impact. Furthermore, we consider here a broken
power law, with 2 different indexes, corresponding to a 
change of slope for the size distributions of the smallest fragments,
a feature which is supported by experimental and theoretical studies
\citep[e.g.][]{dav90,tang99}.

\subsection{cratering}
\label{sec:appB3}

If $Q\dma{imp}<Q^{*}$, the target is preserved
but eroded by a mass $M\dma{cra}$. 
In most published collision--evolution models, $M\dma{cra}$ is directly
proportional to $E\dma{col}$ through a constant coefficient $\alpha$, often
called excavation coefficient \citep[e.g.][]{green78,sto75,pet93} or
defined as $\alpha=1/Q_c$, where $Q_c$ is the ``crushing energy'' 
\citep[e.g.][]{ws93,ken99}.
Values of $\alpha$ typically range between $10^{-9}$s$^{2}$cm$^{-2}$ for hard material
and $4\times10^{-8}$s$^{2}$cm$^{-2}$ for weakly bonded sand
\citep[e.g.][]{green78,dob84,pet93}. In TAB03 we also followed
this prescription, with an intermediate $\alpha=10^{-8}$s$^{2}$cm$^{-2}$ value.
However, the $M\dma{cra}=\alpha E\dma{col}$ relation is in reality a simplification
of the more general 

\begin{equation}
M\dma{cra}=\alpha' E\dma{col}^{\gamma}
\label{Mcra}
\end{equation}
dependence, with
index $\gamma$ slightly greater than 1 \citep[e.g.][]{gau73,kos01}.
Historically, the simplified $M\dma{cra}=\alpha E\dma{col}$ relation has
been derived by \citet{mar69} who extrapolated experimental results,
obtained mainly by \citet{gau62} on small projectiles, to the much larger sizes
considered in his study \citep[see p.77 of][]{mar69}. It has been later
assumed by \citet{green78} in their milestone numerical
study of planetesimal accretion and in most statistical collisional
evolution models ever since. But one should be aware
that this relation is in principle only valid over a
limited range of object sizes and velocities, typically
$10$m$\la R\la 1$km and $\Delta V \simeq$ 3--5\,km.s$^{-1}$,
and that a study considering size ranges spanning over several orders of magnitudes
should assume the ``real'' dependence in $\alpha' E\dma{col}^{\gamma}$.
One of the most accurate $M\dma{cra}$ perscription is probably given in
Equ.7 of \citet{kos01}, which is an empirical fit of experimental
results obtained by
these authors for mixed ice/basalt bodies as well as by several other
studies for pure silicate or pure ice objects and reads (with the present formalism):

\begin{equation}
M\dma{cra} = V\dma{ice} \left(\frac{V\dma{sil}}{V\dma{ice}}\right)^{f\dma{sil}}
2^{-\gamma} \rho E\dma{col}^{\gamma}
\label{McrV}
\end{equation}
where $f\dma{sil}$ is the proportion of silicates in the target,
$V\dma{sil} = 10^{-8}$cgs, $V\dma{ice} = 6.69 \times 10^{-7}$cgs, and $\gamma=1.23$.
Note that this formula gives substantially lower excavated masses
for small ($<1$cm) targets than those derived with the
$M\dma{cra}=10^{-8} E\dma{col}$ relation taken in TAB03.

Nevertheless, this formula is only valid in the small--scale impact regime,
where $M\dma{cra}<<M_t$, corresponding basically to a grain--hitting--a--wall case.
For larger craters, effects of cratering in finite spheres have to be taken
into account \citep{hol94}. This raises the more general issue
of ``connecting'' the cratering prescription to the fragmentation one. 
Some collision--evolution models assume an abrupt fragmentation/cratering transition,
where the maximum possible value of $M\dma{cra}/M\dma{t}$ just below
the fragmentation threshold is $\simeq 0.1$ \citep[e.g.][]{pet93,theb03},
thus implicitly leading to a sharp drop from $(1-M\dma{lf})=0.5 M_t$
to $M\dma{cra}=0.1 M_t$. Experiments seem however to show that there is no
sharp transition around the $M\dma{lf}=0.5 M_t$ value \citep{dav90,hou91},
so that the transition between the fragmentation and cratering
regimes should be more or less progressive. 
Such a smooth transition is assumed for our present model, where
we consider 3 cases. For small--scale craters, we take:

\begin{equation}
M\dma{cra} = \alpha' E\dma{col}^{\gamma} {\rm\,\,\,\,\,\, if \,\,\,}   Q\dma{imp}<0.01 Q^{*}
\end{equation}
with $\alpha'$ given by Equ.\ref{McrV}.
For the large--scale regime just below the fragmentation threshold, we
follow \citet{wya02} and assume:

\begin{equation}
M\dma{cra} = 0.5 M_t\,\left(\frac{Q\dma{imp}}{Q^{*}}\right)
{\rm\,\,\,\,\,\,\,\, if \,\,\,\,} 0.2Q^{*}<Q\dma{imp}<Q^{*}
\end{equation}
which is in agreement with the experiment results
displayed in Fig.5 of \citet{hou91}. 
Between these two modes, we assume a smooth transition given by:

\begin{equation}
M\dma{cra} = K\,E\dma{col}
{\rm\,\,\,\,\,\,\,\, if \,\,\,\,} 0.01Q^{*}<Q\dma{imp}<0.2Q^{*}
\end{equation}
with
\begin{eqnarray}
log(K) = log(K\dma{ls}) \,\,\,\, -
\nonumber \\
\frac{log(\frac{Q\dma{imp}}{Q^{*}})-log(0.2)}{log(0.2)-log(0.01)}
\left(log(K\dma{ss})-log(K\dma{ls})\right)
\end{eqnarray}
where
\begin{equation}
K\dma{ss} = \alpha' E\dma{col}^{\gamma-1}{\rm \,\,\,\,\,\,;\,\,\,\,}
K\dma{ls} =  0.5 \left(\frac{1}{Q^{*}}\right)
\end{equation}

In a similar way to what was assumed for the $Q^{*}$ parameter, we assume that
$M\dma{cra(ice)}=5\,M\dma{cra(sil)}$ as well as $M\dma{cra(ice-sil)}=2\,M\dma{cra(ice)}$
and $M\dma{cra(sil-ice)}=1/2\,M\dma{cra(sil)}$.

The excavated mass $M\dma{cra}$ is then redistributed into fragments following
a single--index size distribution power law (see section 2.5 of TAB03).

\end{appendix}

\clearpage


\begin{thebibliography}{}
%
%
\bibitem[Arakawa(1999)]{ara99} Arakawa, M., 1999, Icarus, 142, 34
%
\bibitem[Ardila et al.(2004)]{ard04} Ardila, D. R.; Golimowski, D. A.; Krist, J. E.;
Clampin, M.; Williams, J. P.; Blakeslee, J. P.; Ford, H. C.; Hartig, G. F.;
Illingworth, G. D., 2004, ApJ, 617, L147
%
\bibitem[Artymowicz(1997)]{arty97} Artymowicz P., 1997, Ann. Rev.
Earth Planet. Sci. 25, 175
%
\bibitem[Augereau et al.(1999)]{aug99} Augereau, J.~C., 
Lagrange, A.~M., Mouillet, D., Papaloizou, J.~C.~B., \& Grorod, P.~A.\ 
1999, \aap, 348, 557 
%
\bibitem[Augereau et al.(2001)]{aug01} Augereau, J.C., Nelson, R.P.,
Lagrange, A.M., Papaloizou, J.C.B., Mouillet, D., 2001, A\&A 370, 447
%
\bibitem[Augereau \& Beust(2006)]{aug06} Augereau, J.-C., Beust, H., 2006,
A\&A, 455, 987
%
\bibitem[Becklin(2006)]{bec06} Becklin, E.~E.\ 2006, 36th 
COSPAR Scientific Assembly, 36, 672 
%
\bibitem[Benz \& Asphaug(1999)]{benz99} Benz, W., Asphaug, E., 1999,
Icarus, 142, 5
%
\bibitem[Burchell et al.(2005)]{bur05} Burchell, M., Leliwa-Kopystynski, J.,
Akawara, M., 2005, Icarus, 179, 274
%
\bibitem[Campo Bagatin et al.(1994)]{bag94} Campo Bagatin, A., Cellino, A.,
Davis, D., Farinella, P., Paolicchi, 1994, Planet. Space Sci., 42, 1079
%
\bibitem[Casey(2006)]{cas06} Casey, S.~C.\ 2006, \procspie, Vol. 6267
%
\bibitem[Davis \& Ryan(1990)]{dav90} Davis, D., Ryan, E., 1990,
Icarus, 83, 156
%
\bibitem[Dent et al.(2000)]{den00} Dent, W.~R.~F., Walker, 
H.~J., Holland, W.~S., \& Greaves, J.~S.\ 2000, \mnras, 314, 702 
%
\bibitem[Dobrovolskis \& Burns(1984)]{dob84} Dobrovolskis, A., Burns, J.A.,
1984, Icarus, 57, 464
%
\bibitem[Dohnanyi(1969)]{dohn69} Dohnanyi J.S., 1969, JGR 74, 2531
%
\bibitem[Dominik \& Decin(2003)]{dom03} Dominik, C.; Decin, G., 2003,
ApJ, 598, 626
%
\bibitem[Draine(2003)]{dra03} Draine, B.~T.\ 2003, \apj, 598, 1026 
%
\bibitem[Draine(2006)]{dra06} Draine, B.~T.\ 2006, \apj, 636, 1114 
%
\bibitem[Durda et al.(1998)]{dur98} Durda, D. D.; Greenberg, R.; Jedicke, R.,
1998, Icarus, 135, 431
%
\bibitem[Gault et al.(1962)]{gau62} Gault, D.E:, Shoemaker, E.M., Moore, H.J.,
1962, NASA TN D-1767
%
\bibitem[Gault(1973)]{gau73} Gault, D.E:, 1973, Moon, 6, 32
%
\bibitem[Golimowski et al.(2006)]{gol06} Golimowski, D.~A., 
et al.\ 2006, \aj, 131, 3109 
%
\bibitem[Greaves et al.(2004)]{gre04} Greaves, J.S., Wyatt, M.C., Holland, W.S.,
Dent, W.R.F. 2004, MNRAS, 351, L54
%
\bibitem[Greaves(2005)]{grea05} Greaves, J.L., 2005, Science, 307, 68
%
\bibitem[Greenberg et al.(1978)]{green78} Greenberg, R.; Hartmann, W. K.;
Chapman, C. R.; Wacker, J. F., 1978, Icarus, 35, 1
%
\bibitem[Grigorieva et al.(2007)]{grig07} Grigorieva, A., Artymowicz, P.,
Th\'ebault, P., 2007, A\&A, 461, 537
%
\bibitem[Henyey \& Greenstein(1941)]{hen41} Henyey, L.G., Greenstein, J.L.,
1941, ApJ, 93, 70
%
\bibitem[Hauschildt et al.(1999)]{hau99} Hauschildt, P.~H., 
Allard, F., \& Baron, E.\ 1999, \apj, 512, 377 
%
\bibitem[Hayashi(1981)]{haya81} Hayashi, C., 1981,PthPS 70, 35  
%
\bibitem[Holland et al.(2006)]{hol06} Holland, W., et al.\ 
2006, \procspie, 6275,  
%
\bibitem[Holsapple(1994)]{hol94} Holsapple, K., 1994, Planet. Space Sci.,
42, 1067
%
\bibitem[Housen \& Holsapple(1990)]{hh90} Housen, K., Holsapple, K., 1990,
Icarus, 84, 226
%
\bibitem[Housen et al.(1991)]{hou91} Housen, K., Schmidt, R.M., Holsapple, K.,
1991, Icarus, 94, 180
%
\bibitem[Kalas \&\ Jewitt(1995)]{kal95} Kalas P., Jewitt D.,
1995, AJ 110, 794
%
\bibitem[Kalas et al.(2006)]{kal06} Kalas, P., Graham, J.~R., 
Clampin, M.~C., \& Fitzgerald, M.~P.\ 2006, \apjl, 637, L57 
%
\bibitem[Kalas et al.(2007)]{kal07} Kalas, P., Fitzgerald, 
M.~P., \& Graham, J.~R.\ 2007, ArXiv e-prints, 704, arXiv:0704.0645 
%
\bibitem[Kenyon \& Luu(1999)]{ken99} Kenyon, S. J.; Luu, Jane X.,
1999, AJ, 118, 1101
%
\bibitem[Kenyon \& Bromley(2002)]{ken02} Kenyon, S. J.;
Bromley, Benjamin C., 2002, AJ, 123, 1757
%
\bibitem[Kenyon \& Bromley(2004)]{ken04} Kenyon, S. J.;
Bromley, Benjamin C., 2004, ApJ, 602, L133
%
\bibitem[Koschny \& Gr\"un(2001)]{kos01} Koschny, D., Gr\"un, E., 2001, Icarus 154, 391
%
\bibitem[Krist et al.(2005)]{kri05} Krist, J.~E., et al.\ 2005, \aj, 129, 1008 
%
\bibitem[Krivov et al.(2000)]{kriv00} Krivov, A., Mann, I.; Krivova, N. A., 2000,
A\&A, 362, 1127
%
\bibitem[Krivov et al.(2005)]{kriv05} Krivov, A., Sremcevic, M., Spahn, F., 2005,
Icarus, 174, 105
%
\bibitem[Krivov et al.(2006)]{kriv06} Krivov, A., Lohne, T., Sremcevic, M., 2006,
A\&A, 455, 509
%
\bibitem[Krivova et al.(2000)]{kriva00} Krivova, N. A.; Krivov, A. V.; Mann, I.,
2000, ApJ, 539, 424
%
\bibitem[Lagrange et al.(2000)]{lag00} Lagrange, A.-M., Backman,
D.~E., \& Artymowicz, P.\, 2000, in {\it Protostars and Planets IV},
the Univ. of Arizona Press, Tucson, 639
%
\bibitem[Li \&\ Greenberg(1998)]{ligre98} Li, A., Greenberg, M., 1998,
A\&A 331, 291
%
\bibitem[Lissauer(1993)]{lis93} Lissauer J., 1993, ARA\&A, 31, 129
%
\bibitem[Lissauer \&\ Stewart(1993)]{lisste93} Lissauer J., Stewart G.,
1993, in {\it Protostars and Planets III}, the Univ. of Arizona Press,
Tucson, 1061
%
\bibitem[Marcus(1969)]{mar69} Marcus, A.J., 1969, Icarus, 11, 76
%
\bibitem[Meyer et al.(2006)]{mey06} Meyer, M.~R., Backman, 
D.~E., Weinberger, A.~J., \& Wyatt, M.~C.\ 2006, in
Protostars and Planets V, Edited by B. Reipurth, D. Jewitt, and K. Keil
University of Arizona Press, Tucson (astro-ph/0606399) 
%
\bibitem[Nakano(1990)]{nak90} Nakano, T.\ 1990, \apjl, 355, 
L43 
%
\bibitem[Paolicchi et al.(1996)]{paol96} Paolicchi, P., Verlicchi, A., Cellino, A.,
1996, Icarus, 121, 126
%
\bibitem[Petit \&\ Farinella(1993)]{pet93} Petit J.-M., Farinella P.,
1993, Celest. Mech. Dynam. Astron. 57, 1
%
\bibitem[Pilbratt(2005)]{pil05} Pilbratt, G.~L.\ 2005, The 
Dusty and Molecular Universe: A Prelude to Herschel and ALMA, 3 
%
\bibitem[Poglitsch et al.(2006)]{pog06} Poglitsch, A., et 
al.\ 2006, 36th COSPAR Scientific Assembly, 36, 215 
%
\bibitem[Schneider et al.(2005)]{sch05} Schneider, G., 
Silverstone, M.~D., \& Hines, D.~C.\ 2005, \apjl, 629, L117 
%
\bibitem[Schneider et al.(2006)]{sch06} Schneider, G., et 
al.\ 2006, \apj, 650, 414 
%
\bibitem[Smith \&\ Terrile(1984)]{smi84} Smith B., Terrile R., 1984,
Sci 226, 1421
%
\bibitem[St\"offler et al.(1975)]{sto75} St\"offler, D., D\"uren, J., Kn\"olker, R.,
Hische, R., Bischoff, A. 1975, Geophys.Res.Lett., 18, 285
%
\bibitem[Strubbe \& Chiang(2006)]{stru06} Strubbe, L.E., Chiang, E.I.,
2006, ApJ, 648, 652
%
\bibitem[Su et al.(2005)]{su05}
Su, K. Y. L.; Rieke, G. H.; Misselt, K. A.; Stansberry, J. A.; Moro-Martin, A.;
Stapelfeldt, K. R.; Werner, M. W.; Trilling, D. E.; Bendo, G. J.; Gordon, K. D.;
Hines, D. C.; Wyatt, M. C.; Holland, W. S.; Marengo, M.; Megeath, S. T.;Fazio, G. G.
2005, ApJ, 628, 487
%
\bibitem[Tanga et al.(1999)]{tang99} Tanga, P., Cellino, A., Michel, P., 
Zappal\`a, V., Paolicchi, P., Dell'Oro, A., 1999, Icarus, 141, 65
%
\bibitem[Th\'ebault \& Augereau (2005)]{theb05} Th\'ebault, P., Augereau, J. C.,
2005, A\&A, 437, 141
%
\bibitem[Th\'ebault et al.(2003)] {theb03} Th\'ebault, P., Augereau, J.-C.,
Beust, H. 2003, A\&A, 408, 775
%
\bibitem[Thommes et al.(2003)]{thom03} Thommes, E. W.; Duncan, M. J.; Levison, H. F.,
2003, Icarus, 161, 431
%
\bibitem[Vidal-Madjar et al.(1994)]{vid94} Vidal-Madjar A.,
Lagrange-Henri A.-M., Feldman P.D., et al., 1994, A\&A 290, 245
%
\bibitem[Wetherill \& Stewart(1993)]{ws93} Wetherill, G.W., Stewart, G.R., 1993,
Icarus, 190.
%
\bibitem[Wyatt \& Dent (2002)]{wya02} Wyatt, M.C.~\& Dent,
W.R.F., 2002, MNRAS, 334, 589
%


\end{thebibliography}
\end{document}